\documentclass[twocolumn,showpacs,preprintnumbers,amsmath,amssymb,aps,pre]{revtex4}

\usepackage{graphicx,color}

\begin{document}

\preprint{}

\title{Statistical mechanics of two-dimensional Euler flows and minimum enstrophy states}

 \author{A. Naso, P.H. Chavanis and B. Dubrulle}

\affiliation{
$^1$ Laboratoire de Physique, Ecole Normale Sup\'erieure de Lyon and CNRS (UMR 5672), 46 all\'ee d'Italie, 69007 Lyon, France\\
$^2$ Laboratoire de Physique Th\'eorique (IRSAMC), CNRS and UPS, Universit\'e de Toulouse, F-31062 Toulouse, France\\
$^3$ SPEC/IRAMIS/CEA Saclay, and CNRS (URA 2464), 91191 Gif-sur-Yvette Cedex, France\\
\email{aurore.naso@ens-lyon.fr, chavanis@irsamc.ups-tlse.fr, berengere.dubrulle@cea.fr}
}

   \date{To be included later }

   \begin{abstract}
A simplified thermodynamic approach of the incompressible
   2D Euler equation is considered based on the conservation of
   energy, circulation and microscopic enstrophy.  Statistical
   equilibrium states are obtained by maximizing the
   Miller-Robert-Sommeria (MRS) entropy under these sole
   constraints. We assume that these constraints are selected by
   properties of forcing and dissipation. We find that the vorticity
   fluctuations are Gaussian while the mean flow is characterized by a
   linear $\overline{\omega}-\psi$ relationship. Furthermore, we prove
   that the maximization of entropy at fixed energy, circulation and
   microscopic enstrophy is equivalent to the minimization of
   macroscopic enstrophy at fixed energy and circulation.  This
   provides a justification of the minimum enstrophy principle from
   statistical mechanics when only the microscopic enstrophy is
   conserved among the infinite class of Casimir
   constraints. Relaxation equations towards the
   statistical equilibrium state are derived. These equations can serve
   as numerical algorithms to determine maximum entropy or minimum
   enstrophy states. We use these relaxation equations to study
   geometry induced phase transitions in rectangular domains. In
   particular, we illustrate with the relaxation equations the
   transition between monopoles and dipoles predicted by Chavanis \&
   Sommeria [J. Fluid. Mech. {\bf 314}, 267 (1996)]. We take into
   account stable as well as metastable states and show that
   metastable states are robust and have negative specific heats. This
   is the first evidence of negative specific heats in that
   context. We also argue that saddle points of entropy can be
   long-lived and play a role in the dynamics because the system may
   not spontaneously generate the perturbations that destabilize them.
\end{abstract}

   \maketitle
%

\section{Introduction} \label{intro}

Two-dimensional turbulence has the striking property of organizing
spontaneously into large-scale
coherent structures. These coherent structures correspond to jets and
vortices in geophysical and astrophysical flows \cite{flierl,marcus}.
They can be reproduced in numerical simulations
\cite{williams} and laboratory experiments
\cite{tabeling,cvh} in either forced or unforced situations.
These coherent structures involve the presence of a mean flow and
fluctuations around it. The mean flow turns out to be a steady state
of the pure 2D Euler equations (without forcing and dissipation) at
some coarse-grained scale.  In several cases, the steady state is
characterized by a linear relationship between vorticity $\omega$ (or
potential vorticity $q=\omega+h$ in the presence of a topography $h$)
and stream function $\psi$. For example, the case of a linear $q-\psi$
relationship was considered early by Fofonoff (1954) \cite{fofonoff}
as a simple model of oceanic circulation. These ``Fofonoff flows''
were found to emerge naturally from random initial conditions in
numerical experiments of forced and unforced 2D turbulence
\cite{veronis,griffa,cummins,wang,kazantsev}.  However, these results
are not expected to be general. There exists many
other cases in 2D turbulence where the $q-\psi$ relationship is not
linear.  The understanding and prediction of these quasi stationary
states (QSS), in forced and unforced situations, is still a
challenging problem. Different approaches have been proposed to
describe these QSSs.

In the case of forced flows (for example oceanic flows experiencing a forcing by the wind and a dissipation),  Niiler (1966) \cite{niiler} and Marshall \& Nurser (1986) \cite{marshall}
have proposed that forcing and dissipation could equilibrate each other in average and determine a
QSS that is a steady state of the unforced and inviscid 2D Euler equation.
This QSS is characterized by a
functional relationship $q=f(\psi)$ between potential vorticity and stream function where the function
$f$ is selected by the properties of forcing and dissipation. In particular, these authors discussed
the properties that forcing must possess  to generate  Fofonoff flows.

In the case of unforced flows, a phenomenological approach, called the
minimum enstrophy principle, was proposed by Bretherton \& Haidvogel
(1976) \cite{bretherton}. It is based on the inverse
cascade\footnote{In 3D turbulence, energy cascades towards smaller and
smaller scales where it is dissipated by viscosity. In 2D turbulence,
the energy is transfered to larger and larger distances in an inverse
cascade (explaining its conservation) whereas this is the enstrophy
that is transfered to shorter and shorter scales, and dissipated.}
process of Batchelor (1969)
\cite{batchelor}.  It is argued that, in the presence of a small
viscosity, the (potential) enstrophy decays while the energy is
approximately conserved. This is what Matthaeus \& Montgomery (1980)
\cite{mm} have called selective decay. In that case, we can expect
that the system will relax towards a state that minimizes (potential)
enstrophy at fixed energy. This is mainly a postulate. This principle
leads to a steady state of the 2D Euler equation characterized by a
linear $q-\psi$ relationship.  When applied to geophysical flows, this
principle can give an alternative justification of Fofonoff flows.
The minimum enstrophy principle has been generalized by Leith (1984)
\cite{leith} so as to take into account the conservation of angular
momentum in order to describe isolated vortices. However, this
principle is purely phenomenological and it is difficult to establish
its domain of validity.

A statistical mechanics approach of 2D turbulence has been developed
by Kraichnan (1967,1975) \cite{k1,k2} based on the truncated Euler
equations.  In that case, the dynamics conserves only the energy and
the enstrophy (the other constraints of the Euler equation are lost by
the truncation). Since this system is Liouvillian, we can apply the
methods of statistical mechanics. This is the
so-called energy-enstrophy statistical theory. In the absence of
topography, the truncated statistical mechanics predicts a homogeneous
flow with an equilibrium energy spectrum of the form $E(k)=k/(a+bk^2)$
corresponding to an equipartition distribution. In the presence of a
topography (or $\beta$-effect), this statistical mechanics approach
was generalized by Salmon, Holloway \& Hendershott (1976)
\cite{salmon}.  It leads to a mean flow where the average potential
vorticity $q$ and the average stream function $\psi$ are related to
each other by a linear relationship.  When applied to geophysical
flows, this approach provides a justification of Fofonoff flows from
statistical mechanics. However, the fact that the truncated
statistical mechanics breaks the conservation of some integrals of the
2D Euler equations may be considered as a limitation of this approach.

Another statistical mechanics of 2D turbulence,  based on a point vortex
approximation, was initiated by Onsager (1949) \cite{onsager}  and further developed by Joyce \& Montgomery (1973) \cite{jm}  and  Lundgren \& Pointin (1977) \cite{lp} in a mean field approximation. The statistical equilibrium state maximizes the usual Boltzmann entropy (adapted to point vortices)
while conserving the energy and the number of vortices of each species. This statistical mechanics predicts an equilibrium state where the relationship between vorticity and stream function is given by a Boltzmann distribution, or a superposition of Boltzmann distributions (on the different species).
However, the point vortex approximation is a crude approximation of real turbulent flows where the vorticity field is continuous.

A statistical theory of 2D turbulence valid for continuous vorticity
fields has been developed by Miller (1990) \cite{miller} and Robert \&
Sommeria (1991)
\cite{rs}.  This theory takes into account all the constraints of the
2D Euler equation. The statistical equilibrium state maximizes a
mixing entropy while conserving energy, circulation and all the
Casimirs. This leads to equilibrium states with more general mean
flows than in the previous approaches. In particular, the
$\overline{\omega}-\psi$ relationship and the fluctuations around it
are determined by the initial conditions and can take various
shapes. However, some connections with the earlier works can be
found. For example, Robert \& Sommeria (1991) \cite{rs} note that the
results of the point vortex approach can be recovered in the dilute
limit of their statistical theory. On the other hand, Miller (1990)
\cite{miller} notes that for specific initial conditions leading to a
Gaussian vorticity distribution at statistical equilibrium, the mean
flow has a linear $\overline{\omega}-\psi$ relationship similar to
that obtained from the minimum enstrophy
principle. Finally, Chavanis \& Sommeria (1996)
\cite{jfm} consider a limit of strong mixing (or low energy) of the
MRS theory in which $\beta\sigma\psi\ll 1$ and find
some connections between maximum entropy states and minimum enstrophy
states. In that limit, the maximization of entropy at fixed energy,
circulation and Casimirs becomes equivalent to the minimization of the
macroscopic enstrophy $\Gamma_2^{c.g.}=\int\overline{\omega}^2\, d{\bf
r}$ at fixed energy and circulation. This justifies a form of inviscid
minimum enstrophy principle in a well-defined limit of the statistical
theory\footnote{In a different situation, Brands {\it et al.} (1999)
\cite{brands} show that, in case of incomplete relaxation, the
statistical equilibrium state restricted to a ``maximum entropy
bubble'' because of lack of ergodicity \cite{jfm2} can resemble a
minimum enstrophy state. However, this agreement is rather fortuitous
and is not expected to be general.}. This strong mixing limit also
makes a hierarchy among the Casimir constraints. To lowest order in
the expansion $\beta\sigma\psi\ll 1$, leading to a linear
$\overline{\omega}-\psi$ relationship, only the circulation $\Gamma$
and the microscopic enstrophy
$\Gamma_2^{f.g.}=\int\overline{\omega^2}\, d{\bf r}$ are important. To
next orders, leading to nonlinear $\overline{\omega}-\psi$
relationships, higher and higher moments
$\Gamma_{n}^{f.g.}=\int\overline{\omega^n}\, d{\bf r}$ become
relevant.

Recently, an alternative statistical theory has been proposed by
Ellis, Haven \& Turkington (2002) \cite{eht} and further discussed by
Chavanis (2005,2008) \cite{physicaD,aussois} and
Chavanis {\it et al.} (2010) \cite{cnd}. These authors argue that, in real
situations where the flows are forced and dissipated at small scales,
the conservation of all the constraints of the 2D Euler equation is
abusive. They propose to keep only the robust constraints (energy and
circulation) and treat the fragile constraints canonically.  This
amounts to prescribing a prior vorticity distribution instead of the
Casimirs. This can be viewed as a grand microcanonical version of the
Miller-Robert-Sommeria (MRS) theory. In the Ellis-Haven-Turkington
(EHT) approach, the statistical equilibrium state maximizes a relative
entropy (determined by the prior) while conserving energy and
circulation. The mean flow turns out to maximize a generalized entropy
determined by the prior at fixed circulation and energy. For a
Gaussian prior, the generalized entropy is proportional to minus the
macroscopic enstrophy. This justifies a minimum enstrophy principle
from statistical mechanics when the constraints are treated
canonically and the prior is Gaussian.  Furthermore, this approach
allows to go beyond the minimum enstrophy principle by considering
more complicated priors.

We can give another interpretation of the EHT approach. Indeed, Bouchet  (2008) \cite{bouchet} notes that the EHT approach provides a {\it sufficient}
condition of MRS stability.  Indeed, it is well-known in statistical mechanics and optimization theory
 \cite{ellis} that a solution
of a maximization problem is always a solution of a more constrained dual maximization
 problem. Thus, grand microcanonical stability (EHT) implies microcanonical stability (MRS).
 Therefore, if the mean flow maximizes a
generalized entropy at fixed circulation and energy, then the
corresponding Gibbs state is a MRS equilibrium state. However, the
reciprocal is wrong in case of ensemble inequivalence (between microcanonical and grand microcanonical ensembles) that is generic  for systems with long-range interactions \cite{ellis}. The mean flow associated to a MRS equilibrium
state does  not necessarily maximize a generalized entropy at fixed circulation and
energy.  In this sense, the
minimization of enstrophy at fixed circulation and energy provides a
sufficient, but not necessary, condition of MRS stability. A review of
the connections between these different variational principles has been
given recently by Chavanis (2009) \cite{proc}.

In this paper, we shall complement these different approaches by
considering a sort of intermediate situation between all these
theories. We argue that, in most physical situations, the system is
forced and dissipated at small scales. In some
situations, forcing and dissipation equilibrate each other so that the
system becomes, in average, statistically equivalent to the 2D Euler
equation where forcing and dissipation are ``switched-off''. In
particular, a quasi stationary state (QSS) can form on a relatively
short timescale. This QSS involves a mean flow and fluctuations arount
it. We propose to describe this state in terms of the statistical
mechanics of the 2D Euler equation. However, we indirectly take into
account the effects of forcing and dissipation in the choice of the
constraints. The energy and the circulation must be obviously
conserved. By contrast,  the conservation of all the
Casimir invariants is abusive and it is likely that some Casimir
invariants will be destroyed by the forcing and
the dissipation.  We argue that some Casimir invariants are more
relevant than others and that they will be selected by the properties
of forcing and dissipation. In that case, the statistical equilibrium
state is expected to maximize the MRS entropy with these sole
constraints. It is not our goal here to determine how these
constraints are selected by the properties of forcing and
dissipation. This is clearly a complicated problem that has to be
tackled by other methods. We shall just use a heuristic approach and
consider the situation where these relevant constraints are the
circulation, the energy and the microscopic enstrophy. In other words,
among all the Casimir constraints, we only consider the quadratic
one. We do not claim that this is the most general situation but
simply that it is a case of physical interest. We therefore consider
{\it the maximization of the MRS entropy at fixed circulation, energy
and microscopic enstrophy} and provide a detailed study of this (non
trivial) variational principle.  Our maximization principle is similar
in spirit to the approach of Kraichnan, which only considers the
conservation of energy and enstrophy. However, we remain in
physical space and work with the MRS entropy for the distribution of
vorticity levels while Kraichnan works in Fourier space. Furthermor, we take into account the conservation of circulation.

It may be noted that our approach is partly
motivated by results of experiments carried out in von Karman flows
\cite{monchaux1,monchaux2}. In fact, the present paper, valid for the 2D Euler
equations, prepares the ground before considering the more complicated
(but similar) case of 3D axisymmetric turbulence treated in
\cite{k}. In that later case, we show the equivalence between
maximum entropy states at fixed helicity, angular momentum and
microscopic energy and minimum macroscopic energy states at fixed
helicity and angular momentum. Therefore, our two approaches are
closely related and provide a statistical basis for justifying the
phenomenological minimum enstrophy (2D) and minimum energy (3D
axisymmetric) principles \cite{mm}.

The paper is organized as follows.

In Sec. \ref{sec_ss}, we discuss
some general properties of the steady states of the 2D Euler
equations. In particular, we mention the refined criterion of
nonlinear dynamical stability given by Ellis {\it et al.} \cite{eht}
based on the maximization of a pseudo entropy at fixed circulation and
energy.

In Sec. \ref{sec_ens}, we recall the phenomenological minimum
enstrophy principle based on selective viscous decay.

In Sec. \ref{sec_equivgen}, we discuss the connection between maximum
entropy states and minimum enstrophy states.  In Sec. \ref{sec_basic},
we consider the maximization of the MRS entropy at fixed energy,
circulation and microscopic enstrophy
(energy-enstrophy-circulation statistical mechanics).
This maximization problem leads to a statistical equilibrium state
with Gaussian fluctuations and a mean flow characterized by a linear
$\overline{\omega}-\psi$ relationship. In Sec. \ref{sec_equiv}, we
introduce an equivalent, but simpler, maximization problem based on
the maximization of an entropic functional of the coarse-grained
vorticity at fixed circulation, energy and microscopic enstrophy. This
generalized entropy has some similarities with the Renyi entropy
\cite{renyi,frank}. In Sec. \ref{sec_reduced}, we show {\it the
equivalence between the maximization of entropy at fixed energy,
circulation and microscopic enstrophy and the minimization of
macroscopic enstrophy at fixed energy and circulation}\footnote{This
equivalence is not trivial. For example, in the MRS approach where all
the Casimirs are conserved, there exists particular initial conditions
leading to a Gaussian Gibbs state at equilibrium associated with a
linear $\overline{\omega}-\psi$ relationship \cite{miller}. However,
we cannot conclude that these states are minimum enstrophy states
(they could be just saddle points of enstrophy at fixed circulation
and energy). We can only prove that a minimum of enstrophy (more
generally a maximum of a ``generalized entropy'') at fixed circulation
and energy is MRS thermodynamically stable \cite{bouchet,proc} but
this is not reciprocal: the coarse-grained vorticity field associated
with a MRS equilibrium state is not necessarily a minimum of enstrophy
(or more generally a maximum of a ``generalized entropy'') at fixed
circulation and energy (it could be a saddle point). By contrast, if we keep only the microscopic
enstrophy as a constraint among all the Casimirs, we shall prove here
that the Gibbs state is a maximum entropy state at fixed circulation,
microscopic enstrophy and energy iff the corresponding coarse-grained
vorticity field is a minimum of macroscopic enstrophy state at fixed
circulation and energy.}.  Therefore, our simplified thermodynamic
approach provides a justification of the minimum enstrophy principle
(and Fofonoff flows) from statistical mechanics when only the
microscopic enstrophy is conserved among the infinite class of Casimir
constraints.

We also derive relaxation equations associated with the various
maximization problems mentioned above (for the
reader's convenience, the formal derivation of these relaxation
equations is postponed to Appendix \ref{sec_re}). Associated with the
basic variational problem, we derive a relaxation equation for the
vorticity distribution that increases the MRS entropy at fixed energy,
circulation and microscopic enstrophy. In that case, the vorticity
distribution $\rho({\bf r},\sigma,t)$ progressively becomes Gaussian
and the mean flow $\overline{\omega}({\bf r},t)$ relaxes towards a
steady state characterized by a linear $\overline{\omega}-\psi$
relationship. Associated with the simplified variational problem, we
derive a relaxation equation for the mean flow that increases a
generalized entropy (Renyi-like) while conserving energy, circulation
and microscopic enstrophy. In that case, the vorticity distribution is
always Gaussian, even in the out-of-equilibrium regime, with a uniform
centered variance monotonically increasing with time.  Associated with
the minimum enstrophy principle, we derive a relaxation equation for
the mean flow that dissipates the macroscopic enstrophy while
conserving energy and circulation.  These relaxation equations can
serve as numerical algorithms to determine maximum entropy or minimum
enstrophy states with appropriate constraints.

In Sec. \ref{sec_ptg}, we study minimum enstrophy states at fixed
energy and circulation in rectangular domains. This problem was first
considered by Chavanis \& Sommeria (1996) \cite{jfm} who report
interesting phase transitions between monopoles and dipoles depending
on the geometry of the domain (e.g., the aspect ratio $\tau$ of a
rectangular domain) and on the value of the control parameter
$\Gamma^2/{E}$. In particular, for $\Gamma=0$, they report a
transition from a monopole to a dipole when the aspect ratio becomes
larger than $\tau_c=1.12$. For $\Gamma\neq 0$ and $\tau<\tau_c$, the
maximum entropy state is always a monopole and for $\Gamma\neq 0$ and
$\tau>\tau_c$, the maximum entropy state is a dipole for small
$\Gamma^2/{E}$ and a monopole for large $\Gamma^2/{E}$. They also
studied the metastability of the solutions (local entropy maxima) and
the possible transition between a direct and a reversed monopole. This
work has been followed recently in different directions. Venaille \&
Bouchet (2009) \cite{vb} have investigated in detail the nature of
these phase transitions from the viewpoint of statistical
mechanics. In particular, they showed that the point
$(\Gamma=0,\tau_c=1.12)$ corresponds to a bicritical point separating
a microcanonical first order transition line to two second order
transition lines.  Keetels, Clercx \& van Heijst (2009)
\cite{keetels} have studied minimum enstrophy states in rectangular or
circular domains with boundary conditions taking into account the
effect of viscosity. Finally, Taylor, Borchardt \& Helander (2009)
\cite{taylor} have shown that the two types of solutions appearing in
the study of Chavanis \& Sommeria \cite{jfm} could explain the process
of ``spin-up'' discovered by Clercx, Maassen \& van Heijst (1998)
\cite{spin}. In Sec. \ref{sec_ptg}, we recall and complete the main
results of the approach of Chavanis \& Sommeria \cite{jfm} in order to
facilitate the discussion of the last section.

In Sec. \ref{sec_relax}, we use the relaxation equations derived in
Appendix \ref{sec_re} to illustrate the phase transitions described in
Sec. \ref{sec_ptg}. On the basis of the relaxation equations, we
observe a persistence of unstable states that are saddle points of
entropy. Therefore, we argue that unstable saddle points of entropy
may play a role in the dynamics if the system does not spontaneously
generate the perturbations that can destabilize them. We also follow
an hysteretic cycle as a function of the circulation where the
hysteresis is due to the robustness of metastable states (local
entropy maxima). Finally, we briefly describe the possibility of
transitions between direct and reversed monopoles in the presence of
stochastic forcing.

Throughout this paper, we consider the simple case of incompressible 2D flows without topography. However, the main formalism of the  theory can be generalized straightforwardly to account for a topography (or a $\beta$-effect) by simply replacing the vorticity $\omega$ by the potential vorticity $q=\omega+h$. Some applications will be considered in a companion paper \cite{fof}. We also assume throughout the  paper that the domain is of unit area  $A=1$.

\section{Dynamical stability of steady states of the 2D Euler equation}
\label{sec_ss}

We consider a two-dimensional incompressible and inviscid flow described by the 2D Euler equations
\begin{eqnarray}
\label{ss1}
\frac{\partial\omega}{\partial t}+{\bf u}\cdot\nabla\omega=0, \qquad -\Delta\psi=\omega,
\end{eqnarray}
where $\omega {\bf z}=\nabla\times {\bf u}$ is the vorticity, $\psi$ the stream function and ${\bf u}$ the velocity field  (${\bf z}$ is a unit vector normal to the flow).  The 2D Euler equation admits an infinite number of steady states
 of the form
\begin{eqnarray}
\label{ss2}
\omega=f(\psi),
\end{eqnarray}
where $f$ is an arbitrary function. They are obtained by solving the
differential equation
\begin{eqnarray}
\label{ss3}
-\Delta\psi=f(\psi),
\end{eqnarray}
with $\psi=0$ on the domain boundary.

To determine the dynamical stability of such flows, we can make use of
the conservation laws of the 2D Euler equations. The 2D Euler
equations conserve an infinite number of integral constraints that are
the energy
\begin{eqnarray}
\label{ss4}
E=\int \frac{{\bf u}^2}{2}\, d{\bf r}=\frac{1}{2}\int \omega\psi\, d{\bf r},
\end{eqnarray}
and the Casimirs
\begin{eqnarray}
\label{ss5}
I_h=\int h(\omega)\, d{\bf r},
\end{eqnarray}
where $h$ is an arbitrary function. In particular, all the moments of
the vorticity $\Gamma_n=\int \omega^n\, d{\bf r}$ are conserved. The
first moment $\Gamma=\int\omega\, d{\bf r}$ is the circulation and the
second moment $\Gamma_2=\int\omega^2\, d{\bf r}$ is the enstrophy. Let
us consider a special class of Casimirs of the form
\begin{eqnarray}
\label{ss6}
S=-\int C(\omega)\, d{\bf r},
\end{eqnarray}
where $C$ is a convex function (i.e. $C''\ge 0$). These functionals are called pseudo entropy \cite{proc}. Ellis {\it et al.} \cite{eht} have shown that the maximization problem
\begin{eqnarray}
\label{ss7}
\max_{\omega}\, \lbrace S[\omega] \, | \, E[\omega]=E, \, \Gamma[\omega]=\Gamma\rbrace,
\end{eqnarray}
determines a steady state of the 2D Euler equation that is nonlinearly dynamically stable. This provides a refined criterion of nonlinear dynamical stability. The critical points of (\ref{ss7}) are given by the variational principle
\begin{eqnarray}
\label{ss8}
\delta S-\beta\delta E-\alpha\delta\Gamma=0,
\end{eqnarray}
where $\beta$ and $\alpha$ are Lagrange multipliers. This gives
\begin{eqnarray}
\label{ss9}
C'(\omega)=-\beta\psi-\alpha \Rightarrow \omega=F(\beta\psi+\alpha),
\end{eqnarray}
where $F(x)=(C')^{-1}(-x)$. We note that $\omega'(\psi)=-\beta/C''(\omega)$ so that $\omega(\psi)$ is a monotonic function increasing for $\beta<0$ and decreasing for $\beta>0$. This critical point is a steady state of the 2D Euler equation. On the other hand, it is a (local) maximum of the pseudo entropy at fixed energy and circulation iff
\begin{eqnarray}
\label{ss10}
-\frac{1}{2}\int C''(\omega)(\delta\omega)^2\, d{\bf r}-\frac{1}{2}\beta\int\delta\omega\delta\psi\, d{\bf r}<0,
\end{eqnarray}
for all perturbations $\delta\omega$ that conserve energy and
circulation at first order. In that case, it is formally nonlinearly
dynamically stable with respect to the 2D Euler equations. This
criterion is stronger than the well-known Arnol'd theorems that only
provide sufficient conditions of stability. We note, however, that the
refined criterion (\ref{ss7}) provides itself just a sufficient
condition of nonlinear dynamical stability. An even
more refined criterion of dynamical stability is given by the
Kelvin-Arnol'd principle. A review of the connections
between these different stability criteria has been recently given by Chavanis
\cite{proc}.

\section{The minimum enstrophy principle}
\label{sec_ens}

Let us  consider the minimization of the enstrophy $\Gamma_2=\int \omega^2\, d{\bf r}$
at fixed circulation and energy
\begin{eqnarray}
\label{ens1}
\min_{\omega}\, \lbrace \Gamma_2[\omega] \, | \, E[\omega]=E, \, \Gamma[\omega]=\Gamma\rbrace.
\end{eqnarray}
The critical points are given by the variational principle
\begin{eqnarray}
\label{ens2}
\delta \Gamma_2+2\beta\delta E+2\alpha\delta\Gamma=0,
\end{eqnarray}
where $2\beta$ and $2\alpha$ are Lagrange multipliers (the factor $2$ has been introduced for compatibility with the results of Sec. \ref{sec_reduced}). This yields
\begin{eqnarray}
\label{ens3}
\omega=-\Delta\psi=-{\beta}\psi-{\alpha}.
\end{eqnarray}
This is a steady state of the 2D Euler equation characterized by a linear $\omega-\psi$ relationship.
On the other hand it is a (local) minimum of enstrophy at fixed energy and circulation iff
\begin{eqnarray}
\label{ens4}
\int (\delta\omega)^2\, d{\bf r}+\beta\int\delta\omega\delta\psi\, d{\bf r}>0,
\end{eqnarray}
for all perturbations $\delta\omega$ that conserve energy and circulation at first order.

There are several interpretations of the minimization principle (\ref{ens1})\footnote{These interpretations can be generalized to any functional of the form $S=-\int C(\omega)\, d{\bf r}$ where $C$ is convex \cite{proc}.}:

(i) The minimum enstrophy principle was introduced in a phenomenological manner from a selective decay principle \cite{bretherton,mm,leith}. Due to a small viscosity, or other source of dissipation, the enstrophy (fragile integral) is dissipated while the energy and the circulation (robust integrals) are relatively well conserved\footnote{If $\nu$ denotes the viscosity, we find from the Navier-Stokes equations that $\dot\Gamma_2=-2\nu\int (\nabla\omega)^2\, d{\bf r}$. When $\nu\rightarrow 0$ and $(\nabla\omega)^2\rightarrow +\infty$, as the flow develops small scales, the product $\nu\int (\nabla\omega)^2\, d{\bf r}$ tends to a strictly positive finite value.  Therefore, the enstrophy decays. By contrast, $\dot E=-\nu\Gamma_2$ tends to zero when $\nu\rightarrow 0$ so that the energy is relatively well conserved.}. It is then argued that the system should reach a minimum enstrophy state at fixed circulation and energy. Note that there is no real justification for this last assumption as discussed in \cite{brands}. The enstrophy could decay without reaching its minimum. Furthermore, the minimum potential enstrophy principle is difficult to justify in terms of viscous effects for the QG equations (see the Appendix of \cite{fof}).

(ii) If we view $\Gamma_2$ as a Casimir of the form (\ref{ss6}), the
minimization principle (\ref{ens1}) is equivalent to the maximization
principle (\ref{ss7}) for the pseudo entropy $S=-\frac{1}{2}\Gamma_2$. In this context, it
determines a particular steady state of the 2D Euler equation that is
nonlinearly dynamically stable according to the refined stability criterion of Ellis {\it et al.} \cite{eht}.
This provides another justification of the minimization problem (\ref{ens1}) in
relation to the inviscid 2D Euler equation.

(iii) For inviscid flows, the microscopic enstrophy $\Gamma_2^{f.g.}=\int \overline{\omega^2}\, d{\bf r}$ is conserved by the 2D Euler equation but the macroscopic enstrophy $\Gamma_2^{c.g.}=\int \overline{\omega}^2\, d{\bf r}$ calculated with the coarse-grained vorticity decreases as enstrophy is lost in the fluctuations. Indeed, by Schwartz inequality: $\Gamma_2^{c.g.}=\int \overline{\omega}^2\, d{\bf r}\le \int \overline{\omega^2}\, d{\bf r}=\Gamma_2^{f.g.}$. By contrast, the energy $E=\frac{1}{2}\int \overline{\omega}\psi\, d{\bf r}$ and the circulation $\Gamma=\int\overline{\omega}\, d{\bf r}$ calculated with the coarse-grained vorticity are approximately conserved. This suggests an inviscid minimum enstrophy principle based on the minimization of macroscopic enstrophy at fixed energy and circulation \cite{jfm}. In this case, selective decay is due to the operation of coarse-graining, not viscosity.

In the following, we shall discuss some connections between the minimum enstrophy principle and the maximum entropy principle.

\section{Connection between maximum entropy states and minimum
enstrophy states}
\label{sec_equivgen}

\subsection{Energy-enstrophy-circulation statistical theory}
\label{sec_basic}

Starting from a generically unstable or unsteady initial condition,
the 2D Euler equations are known to develop a complicated mixing
process leading ultimately to a quasi stationary state (QSS), a vortex
or a jet, on the coarse-grained scale. In order to describe this QSS
and the fluctuations around it, we must introduce a probabilistic
description. Let us introduce the density probability $\rho({\bf
r},\sigma)$ of finding the vorticity level $\omega=\sigma$ at position
${\bf r}$. Then, the local moments of vorticity are
$\overline{\omega^n}=\int \rho\sigma^n\, d{\bf r}$.  In the
statistical mechanics approach of Miller-Robert-Sommeria
\cite{miller,rs}, assuming that the system is strictly described by
the 2D Euler equation (no forcing and no dissipation), the statistical
equilibrium state is expected to maximize the mixing entropy
\begin{equation}
S[\rho] = - \int \rho \ln \rho \, d{\bf r}d\sigma,
\label{es1}
\end{equation}
while conserving all the invariants (energy and Casimirs) of the 2D Euler equation. This forms the standard MRS theory.

In the case of flows that are forced and dissipated at small scales,
one may argue that forcing and dissipation will compensate each other
in average so that the system will again achieve a QSS that is a
stationary solution of the 2D Euler equation. This QSS will be
selected by forcing and dissipation. This fact is vindicated both in
experiments \cite{monchaux1,monchaux2} and numerical simulations
\cite{bousim}. In order to describe the fluctuations around this
state, one needs to go one step further and obtain the vorticity
distribution. An idea is to keep the framework of the statistical
theory but argue that forcing and dissipation will alter the
constraints. More precisely, forcing and dissipation will select some
particular relevant constraints among all the invariants of the ideal
2D Euler equation. These constraints will determine the mean flow and
the fluctuations around it. For example, we argue that there exists
physical situations in which only the conservation of energy,
circulation and microscopic enstrophy are relevant (and of course the
normalization condition). We do not claim that such situations are
universal but simply that they happen in some cases of physical
interest. This seems to be the case for example in some oceanic
situations
\cite{fofonoff,veronis,griffa,cummins,wang,kazantsev,niiler,marshall,bretherton,batchelor,mm,leith,k1,k2,salmon} (see also \cite{df} for recent studies). We shall therefore consider the
maximization problem
\begin{equation}
\max_{\rho}\, \lbrace S[\rho]\, | \, E,\, \Gamma, \, \Gamma_2^{f.g.}, \, \int \rho d\sigma=1  \rbrace,
\label{es2}
\end{equation}
where
\begin{eqnarray}
E=\frac{1}{2}\int\overline{\omega}\psi\, d{\bf r}=\frac{1}{2}\int \psi \rho\sigma \, d{\bf r}d\sigma,
\label{es3}
\end{eqnarray}
\begin{eqnarray}
\Gamma=\int\overline{\omega}\, d{\bf r}=\int  \rho\sigma \, d{\bf r}d\sigma,
\label{es4}
\end{eqnarray}
\begin{eqnarray}
\Gamma_{2}^{f.g.}=\int\overline{\omega^2}\, d{\bf r}=\int  \rho\sigma^2 \, d{\bf r}d\sigma.
\label{es5}
\end{eqnarray}
The last constraint (\ref{es5}) will be called the microscopic (or fine-grained) enstrophy because it takes into account the fluctuations of the vorticity $\omega$. It is different from the macroscopic (or coarse-grained) enstrophy
\begin{eqnarray}
\Gamma_2^{c.g.}=\int \overline{\omega}^2\, d{\bf r}.\label{es13}
\end{eqnarray}
which ignores these fluctuations. We have
\begin{eqnarray}
\Gamma_2^{f.g.}=\int \omega_2\, d{\bf r}+\Gamma_2^{c.g.},\label{es14}
\end{eqnarray}
where $\omega_2\equiv \overline{\omega^2}-\overline{\omega}^2$ is the
local centered variance of the vorticity. The fluctuations of
enstrophy are $\Gamma_2^{fluct}=\int \omega_2\, d{\bf r}$.  In our
terminology, the enstrophy will be called a {\it fragile constraint}
because it cannot be expressed in terms of the coarse-grained field
since $\overline{\omega^2}\neq
\overline{\omega}^2$. While the microscopic enstrophy
$\Gamma_{2}^{f.g.}$ is conserved, the macroscopic enstrophy
$\Gamma_{2}^{c.g.}$ is not conserved and decays. By
contrast, the energy (\ref{es3}) and the circulation (\ref{es4}) will
be called {\it robust constraints} because they can be expressed in
terms of the coarse-grained fields. We shall come back to this
important distinction in Sec. \ref{sec_reduced}.

The maximization problem (\ref{es2}) is what we shall call
``the energy-enstrophy-circulation statistical
theory", or simply, ``the statistical theory" in this paper.  We note
that a solution of (\ref{es2}) is always a MRS statistical equilibrium
state, but the reciprocal is wrong in case of ensemble inequivalence
because we have relaxed some constraints (the other Casimirs). Here,
we keep the robust constraints $E$ and $\Gamma$ and only {\it one}
fragile constraint $\Gamma_2^{f.g.}$, the quadratic one. We assume
that these constraints are selected by the properties of forcing and
dissipation. As we shall see, this assumption leads to Gaussian
fluctuations and a mean flow characterized by a linear
$\overline{\omega}-\psi$ relationship. In principle, we can obtain
more complex fluctuations and more complex mean flows
(characterized by nonlinear $\overline{\omega}-\psi$
relationships) by keeping more and more fine-grained moments
$\Gamma_{n>1}^{f.g.}$ among the constraints. This can be a practical
way to go beyond the Gaussian approximation. However,
the Gaussian approximation, leading to a linear
$\overline{\omega}-\psi$ relationship, is already an interesting
problem presenting rich bifurcations (because the energy constraint is
nonlinear) \cite{jfm}, so that we shall stick to that situation.

The critical points of (\ref{es2}) are solution of the variational principle
\begin{eqnarray}
\delta S - \beta \delta {E}  -  \alpha \delta \Gamma-  \alpha_2 \delta \Gamma_2^{f.g.} -\int \zeta({\bf r})\delta\left (\int \rho d\sigma\right )\, d{\bf r}= 0,\nonumber\\
\label{es6}
\end{eqnarray}
where $\beta$, $\alpha$, $\alpha_2$ and $\zeta({\bf r})$ are Lagrange multipliers. This yields the Gibbs state
\begin{equation}
\rho({\bf r},\sigma) = \frac{1}{Z({\bf r})} e^{-\alpha_2\sigma^2}e^{-(\beta\psi+\alpha)\sigma},
\label{es7}
\end{equation}
where the ``partition function'' is determined via the normalization condition
\begin{equation}
Z({\bf r}) = \int e^{-\alpha_2\sigma^2}e^{-(\beta\psi+\alpha)\sigma} \, d\sigma.\label{es8}
\end{equation}
Therefore, in this approach, the distribution $\rho({\bf r},\sigma)$ of the
fluctuations of vorticity is Gaussian and the  centered variance of the vorticity $\omega_2({\bf r})$ is uniform
\begin{equation}
\omega_2\equiv \overline{\omega^2}-\overline{\omega}^2=\frac{1}{2\alpha_2}\equiv\Omega_2.\label{es9}
\end{equation}
On the other hand, the mean flow is given by
\begin{equation}
\overline{\omega}=-\Omega_2 (\beta\psi+\alpha).\label{es10}
\end{equation}
This is a steady state of the 2D Euler equation characterized by a
linear $\overline{\omega}-\psi$ relationship. Then, the Gibbs state can be rewritten
\begin{equation}
\rho({\bf r},\sigma) = \frac{1}{\sqrt{2\pi\Omega_2}}e^{-\frac{(\sigma-\overline{\omega})^2}{2\Omega_2}}.
\label{es11}
\end{equation}
Since $\omega_2=\Omega_2$ is
uniform at statistical equilibrium, we get
\begin{eqnarray}
\Gamma_2^{f.g.}=\Omega_2+\Gamma_2^{c.g.}.\label{es15}
\end{eqnarray}
Finally, a critical point of (\ref{es2}) is an entropy {\it maximum} at fixed $E$, $\Gamma$ and $\Gamma_2^{f.g.}$ iff
\begin{eqnarray}
\delta^2 J\equiv -\frac{1}{2}\int \frac{(\delta\rho)^2}{\rho}\, d{\bf r}d\sigma
-\frac{1}{2}\beta\int\delta\overline{\omega}\delta\psi\, d{\bf r}<0
\label{es12}
\end{eqnarray}
for all perturbations $\delta\rho$ that conserve energy, circulation, microscopic enstrophy and normalization at first order (the proof is similar to the one given in \cite{proc} for related maximization problems).

{\it Remark:} From Eqs. (\ref{es1}) and (\ref{es11}), we easily find that $S=\frac{1}{2}\ln\Omega_2$. Then, using Eq. (\ref{es15}), we conclude that, at equilibrium, the entropy is given by
\begin{eqnarray}
S=\frac{1}{2}\ln\left (\Gamma_2^{f.g.}-\Gamma_2^{c.g.}\right ).
\label{enteq}
\end{eqnarray}
Therefore, if there exists several local entropy maxima (metastable
states) for the same values of the constraints $E$, $\Gamma$ and
$\Gamma_2^{f.g.}$, the maximum entropy state is the one with the
smallest enstrophy $\Gamma_{2}^{c.g.}$. This is a first result showing
connections between maximum entropy and minimum enstrophy
principles. However, this {\it equilibrium} result does not prove that
the maximization of the entropy functional $S[\rho]$ at fixed $E$,
$\Gamma$ and $\Gamma_2^{f.g.}$ is equivalent to the minimization of
the enstrophy functional $\Gamma_2^{c.g.}[\overline{\omega}]$ at fixed
$E$ and $\Gamma$ (e.g. an entropy maximum could be a saddle point of
enstrophy). This equivalence will be shown in Sec. \ref{sec_reduced}.

{\it Remark:} the maximization problem (\ref{es2}) also arises in the study of Kazantsev {\it et al.} \cite{kazantsev} when the relaxation equations associated with the MRS statistical theory are closed by using a Gaussian approximation (see \cite{cnd} for details).

\subsection{An equivalent but simpler variational principle}
\label{sec_equiv}

The maximization problem (\ref{es2}) is difficult to solve, especially
regarding the stability condition (\ref{es12}), because we have to
deal with a distribution $\rho({\bf r},\sigma)$. We shall introduce
here an equivalent but simpler maximization problem by ``projecting''
the distribution on a smaller subspace. To solve the maximization
problem (\ref{es2}), we can proceed in two steps\footnote{This
``two-steps'' method was used by one of us (PHC) in different contexts
\cite{aussois,assise,proc}.}:

{\it (i) First step:} We first maximize $S$ at fixed $E$, $\Gamma$, $\Gamma_2^{f.g.}$, $\int \rho\, d\sigma=1$ {\it and} a given vorticity profile $\overline{\omega}({\bf r})=\int\rho\sigma\, d\sigma$. Since the specification of $\overline{\omega}({\bf r})$ determines $E$ and $\Gamma$, this is equivalent to maximizing $S$ at fixed $\Gamma_2^{f.g.}$,  $\int \rho\, d\sigma=1$ {and} $\overline{\omega}({\bf r})=\int\rho\sigma\, d\sigma$. Writing the variational problem as
\begin{eqnarray}
\label{ev1}
\delta S-\alpha_2\delta\Gamma_2^{f.g.}-\int\lambda({\bf r})\delta\left (\int\rho\sigma\, d\sigma\right )\, d{\bf r}\nonumber\\
-\int \zeta({\bf r})\delta\left (\int\rho\, d\sigma\right )\, d{\bf r}=0,\qquad
\end{eqnarray}
we obtain
\begin{equation}
\rho_1({\bf r},\sigma) = \frac{1}{\sqrt{2\pi\Omega_2}}e^{-\frac{(\sigma-\overline{\omega})^2}{2\Omega_2}},
\label{ev2}
\end{equation}
with
\begin{eqnarray}
\omega_2 \equiv
\overline{\omega^2}-\overline{\omega}^2=\frac{1}{2\alpha_2}\equiv\Omega_2.
\label{ev3}
\end{eqnarray}
Note that the centered variance of the vorticity $\omega_2({\bf
r})=\Omega_2$ is {\it uniform}. Equation (\ref{ev3}) also implies that
$\alpha_2$ must be positive. We check that $\rho_1$ is a global
entropy maximum with the previous constraints since $\delta^2
S=-\int\frac{(\delta\rho)^2}{2\rho}\, d{\bf r}d\sigma<0$ (the
constraints are linear in $\rho$ so their second variations vanish).
Using the optimal distribution (\ref{ev2}), we can express the
entropy (\ref{es1}) in terms of $\overline{\omega}$
writing $S[\overline{\omega}]\equiv S[\rho_1]$.  After straightforward
calculations we obtain
\begin{equation}
S=\frac{1}{2}\ln\Omega_2,
\label{ev4}
\end{equation}
up to some unimportant constant terms. Note that $\Omega_2$ is determined by the constraint on the microscopic enstrophy which leads to
\begin{eqnarray}
\Omega_2=\Gamma_{2}^{f.g.}-\int\overline{\omega}^2 \, d{\bf r}.
\label{ev5}
\end{eqnarray}
The second term is the macroscopic enstrophy associated with
the coarse-grained flow: $\Gamma_{2}^{c.g.}=\int\overline{\omega}^2 \,
d{\bf r}$.  The relation (\ref{ev5}) can be used to express the
entropy (\ref{ev4}) in terms of $\overline{\omega}$ alone.

{\it (ii) Second  step:} we now have to solve the maximization problem
\begin{equation}
\max_{\overline{\omega}}\lbrace S[\overline{\omega}]\, | \, E, \, \Gamma,\, \Gamma_2^{f.g.}  \rbrace,\label{ev6}
\end{equation}
with
\begin{equation}
S=\frac{1}{2}\ln\left (\Gamma_2^{f.g.}-\int \overline{\omega}^2\, d{\bf r}\right ),
\label{ev7}
\end{equation}
\begin{eqnarray}
E=\frac{1}{2}\int\overline{\omega}\psi\, d{\bf r},
\label{ev8}
\end{eqnarray}
\begin{eqnarray}
\Gamma=\int\overline{\omega}\, d{\bf r}.
\label{ev9}
\end{eqnarray}
The functional (\ref{ev7}) might be called a generalized
entropy. Interestingly, it resembles the Renyi entropy \cite{renyi,frank}.

{\it (iii) Conclusion:} finally, the solution of
(\ref{es2}) is given by Eq. (\ref{ev2}) where $\overline{\omega}$ is
determined by (\ref{ev6}). Therefore, (\ref{es2}) and (\ref{ev6}) are
equivalent but (\ref{ev6}) is easier to solve because it is expressed
in terms of $\overline{\omega}({\bf r})$ while (\ref{es2}) is expressed in
terms of $\rho({\bf r},\sigma)$.

Up to second order, the variations of entropy (\ref{ev7}) are given by
\begin{eqnarray}
\Delta S=-\frac{1}{\Omega_2}\biggl (\int \overline{\omega}\delta\overline{\omega}\, d{\bf r}+\frac{1}{2}\int (\delta\overline{\omega})^2\, d{\bf r}\biggr )\nonumber\\
-\frac{1}{(\Omega_2)^2}\biggl (\int \overline{\omega}\delta\overline{\omega}\, d{\bf r} \biggr )^2,
\label{ev10}
\end{eqnarray}
where $\Omega_2$ is given by Eq. (\ref{ev5}). Considering the first order variations of the entropy, the critical points of
(\ref{ev6}) are determined by the variational problem
\begin{equation}
\delta S-\beta\delta E-\alpha\delta \Gamma=0.
\label{ev11}
\end{equation}
This yields
\begin{equation}\label{ev12}
\overline{\omega}=-\Omega_2 (\beta {\psi} + \alpha).
\end{equation}
We recover Eq. (\ref{es10}) for the mean flow. Combined with
Eq. (\ref{ev2}), we recover the Gibbs state (\ref{es7}). Considering
now the second order variations of the entropy (\ref{ev10}), we find that a
critical point of (\ref{ev6}) is a maximum of entropy at fixed energy,
circulation and microscopic enstrophy iff
\begin{eqnarray}
-\frac{1}{2\Omega_2}\int (\delta\overline{\omega})^2\, d{\bf r}-\frac{\beta}{2}\int\delta\overline{\omega}\delta\psi\, d{\bf r}\nonumber\\
-\frac{1}{(\Omega_2)^2}\biggl (\int \overline{\omega}\delta\overline{\omega}\, d{\bf r} \biggr )^2
<0,
\label{ev13}
\end{eqnarray}
for all perturbations $\delta\overline{\omega}$ that conserve
circulation and energy at first order (the conservation of microscopic
enstrophy is automatically taken into account in our
formulation). This stability condition is equivalent to Eq. (\ref{es12})
but much simpler because it depends only on the perturbation
$\delta\overline{\omega}$ instead of the perturbation of the full
distribution $\delta\rho$. In fact, the stability condition (\ref{ev13}) can be simplified further. Indeed, using Eq. (\ref{ev12}), we find that the term in parenthesis can be written
\begin{eqnarray}
\int \overline{\omega}\delta\overline{\omega}\, d{\bf r}=-\Omega_2 \int (\beta {\psi} + \alpha)\delta\overline{\omega}\, d{\bf r},
\label{ev13b}
\end{eqnarray}
and it vanishes since the energy and the circulation are conserved at first order so that $\delta E=\int \psi\delta{\omega}\, d{\bf r}=0$ and $\delta \Gamma=\int \delta{\omega}\, d{\bf r}=0$. Therefore, a
critical point of (\ref{ev6}) is a maximum of entropy at fixed energy,
circulation and microscopic enstrophy iff
\begin{eqnarray}
-\frac{1}{2\Omega_2}\int (\delta\overline{\omega})^2\, d{\bf r}-\frac{\beta}{2}\int\delta\overline{\omega}\delta\psi\, d{\bf r}<0,
\label{ev13c}
\end{eqnarray}
for all perturbations $\delta\overline{\omega}$ that conserve
circulation and energy at first order. In fact, this stability condition can be obtained more rapidly if we remark that the maximization problem (\ref{ev6}) is equivalent to the minimization of the macroscopic enstrophy at fixed energy and circulation (see Sec. \ref{sec_reduced}).

\subsection{Equivalence with the minimum enstrophy principle}
\label{sec_reduced}

Since $\ln(x)$ is a monotonically increasing function, it is clear that the maximization problem (\ref{ev6}) is equivalent to
\begin{equation}
\max_{\overline{\omega}}\lbrace S[\overline{\omega}]\, | \, E,\, \Gamma  \rbrace,\label{red14}
\end{equation}
with
\begin{equation}
S=-\frac{1}{2} \Gamma_{2}^{c.g.},
\label{red10}
\end{equation}
\begin{equation}
 \Gamma_{2}^{c.g.}=\int\overline{\omega}^2\, d{\bf r},
\label{red11}
\end{equation}
\begin{equation}
E=\frac{1}{2}\int \overline{\omega}\psi\, d{\bf r},
\label{red12}
\end{equation}
\begin{equation}
\Gamma=\int \overline{\omega}\, d{\bf r}.
\label{red13}
\end{equation}
The functional $S$ of the coarse-grained vorticity $\overline{\omega}$
is called a ``generalized entropy''. It is proportional to the
opposite of the coarse-grained enstrophy. We have the equivalences
\begin{equation}
 (\ref{red14}) \Leftrightarrow  (\ref{ev6})\Leftrightarrow (\ref{es2}).\label{red16}
\end{equation}
Therefore, the maximization of MRS entropy at fixed energy, circulation
and microscopic enstrophy is equivalent to the minimization of
macroscopic enstrophy at fixed energy and circulation. The solution of
(\ref{es2}) is given by Eq. (\ref{ev2}) where $\overline{\omega}$ is
determined by (\ref{red14}) and $\Omega_2$ by
Eq. (\ref{ev5}). Therefore, (\ref{es2}) and (\ref{red14}) are
equivalent but (\ref{red14}) is easier to solve because it is
expressed in terms of $\overline{\omega}$ while (\ref{es2}) is
expressed in terms of $\rho$. This provides a justification of the
coarse-grained minimum enstrophy principle in terms of statistical
mechanics when only the microscopic enstrophy is conserved among the
Casimirs.  Note that, according to (\ref{ss7}), the principle
(\ref{red14}) also assures that the mean flow associated with the
statistical equilibrium state (\ref{es2}) is nonlinearly dynamically
stable with respect to the 2D Euler equation.

The critical points of (\ref{red14}) are given by the variational problem
\begin{equation}
\delta S-\beta\delta E-\alpha\delta \Gamma=0.
\label{red17}
\end{equation}
This yields
\begin{equation}
\overline{\omega}=-\beta\psi-\alpha.\label{red18}
\end{equation}
This returns Eq. (\ref{es10}) for the mean flow (up to a trivial redefinition of $\beta$ and $\alpha$). Together with Eq. (\ref{ev2}), this returns the Gibbs state (\ref{es7}). On the other hand, this state is a maximum of $S$ at fixed $E$ and $\Gamma$  iff
\begin{eqnarray}
-\frac{1}{2}\int (\delta\overline{\omega})^2\, d{\bf r}
-\frac{\beta}{2}\int\delta\overline{\omega}\delta\psi\, d{\bf r}<0,
\label{red19}
\end{eqnarray}
for all perturbations $\delta\overline{\omega}$ that conserve
circulation and energy at first order.  This is equivalent to the criterion (\ref{ev13c}) as it should.

We have thus shown the equivalence between the maximization of MRS entropy at fixed energy, circulation and fine-grained enstrophy with the minimization of coarse-grained enstrophy at fixed energy and circulation. This equivalence has been shown here for global maximization. In Appendix \ref{sec_eqloc}, we prove the equivalence for local maximization by showing that the stability criteria (\ref{es12}) and (\ref{red19}) are equivalent.

\subsection{Equivalence with a grand microcanonical ensemble}
\label{sec_grandmicro}

In the basic maximization problem (\ref{es2}), the fine-grained
enstrophy is treated as a constraint. Let us introduce a grand
microcanonical ensemble by making a Legendre transform of the entropy
with respect to this fragile constraint $\Gamma_2^{f.g.}$
\cite{proc}. We thus introduce the functional $S_g=S-\alpha_2
\Gamma_2^{f.g.}$ and the maximization problem
\begin{equation}
\max_{\rho}\lbrace S_g[\rho]\, | \, E,\, \Gamma, \, \int \rho d\sigma=1  \rbrace.\label{gm1}
\end{equation}
A solution of (\ref{gm1}) is always a solution of the more constrained
dual problem (\ref{es2}) but the reciprocal is wrong in case of
``ensemble inequivalence'' \cite{proc}. In the present case, however,
we shall show that the microcanonical ensemble (\ref{es2}) and the
grand microcanonical ensemble (\ref{gm1}) are equivalent. This is
because only a quadratic constraint (enstrophy) is involved.

To solve the maximization problem (\ref{gm1}) we can proceed in two
steps. We first maximize $S_g$ at fixed $E$, $\Gamma$, $\int
\rho\, d\sigma=1$ {\it and} $\overline{\omega}({\bf r})=\int\rho\sigma\,
d\sigma$.  This is equivalent to maximizing $S_g$ at fixed $\int
\rho\, d\sigma=1$ and $\overline{\omega}({\bf r})=\int\rho\sigma\,
d\sigma$, and this leads to the optimal distribution (\ref{ev2}) where
$\Omega_2=1/(2\alpha_2)$ is now fixed. This is clearly the global
maximum of $S_g$ with the previous constraints. Using this optimal
distribution, we can now express the functional $S_g$ in terms of
$\overline{\omega}$ by writing
$S[\overline{\omega}]=S_g[\rho_1]$. After straightforward
calculations, we obtain
\begin{equation}
S_g=-\frac{1}{2\Omega_2} \Gamma_2^{c.g.},
\label{gm2}
\end{equation}
up to some constant terms (recall that $\Omega_2$
is a fixed parameter in the present situation). In the
second step, we have to solve the maximization problem
\begin{equation}
\max_{\overline{\omega}}\lbrace S[\overline{\omega}]\, | \, E,\, \Gamma  \rbrace.
\label{gm3}
\end{equation}
Finally, the solution of (\ref{gm1}) is given by Eq. (\ref{ev2}) where
$\overline{\omega}$ is determined by (\ref{gm3}). Therefore, the
variational principle (\ref{gm1}) is equivalent to (\ref{gm3}). That
this is true also for local maximization is shown in Appendix B of
\cite{cnd} (in a more general situation).  On the other hand,
since $\Omega_2>0$, the maximization problem (\ref{gm3}) is equivalent
to (\ref{red14}). Since we have proven previously that (\ref{red14})
is equivalent to the microcanonical variational principle (\ref{es2}),
we conclude that (\ref{es2}) and (\ref{gm1}) are equivalent.

{\it Remark:} the grand microcanonical ensemble (\ref{gm1}) corresponds to the EHT approach with a Gaussian prior \cite{eht,aussois,cnd}.

\subsection{Connection between different variational principles}
\label{sec_conn}

Let us finally discuss the relationship between our approach, Naso-Chavanis-Dubrulle (NCD), and the ones proposed by Miller-Robert-Sommeria (MRS) and Ellis-Haven-Turkington (EHT). To that purpose, we shall make the connection between the variational principles  \cite{proc}:

\begin{equation}
({\rm MRS}): \qquad \max_\rho \lbrace S[\rho] \, | \, E,\Gamma,\Gamma_{n>1}^{f.g.},\int\rho d\sigma=1 \rbrace,
\end{equation}
\begin{equation}
({\rm EHT}): \max_\rho \lbrace S_\chi[\rho] \, | \, E,\Gamma,\int\rho d\sigma=1 \rbrace,
\end{equation}
\begin{equation}
({\rm NCD}): \max_\rho \lbrace S[\rho] \, | \, E,\Gamma,\Gamma_2^{f.g.},\int\rho d\sigma=1 \rbrace.
\end{equation}
\begin{equation}
({\rm MaxS}): \max_{\overline{\omega}} \lbrace S[\overline{\omega}] \, | \, E,\Gamma \rbrace.
\end{equation} 
\begin{equation}
({\rm Min\Gamma_2}): \min_{\overline{\omega}} \lbrace \Gamma_2^{c.g.}[\overline{\omega}] \, | \, E,\Gamma \rbrace,
\end{equation}
where the functionals are
\begin{equation}
S[\rho]=-\int \rho({\bf r},\sigma)\ln\rho({\bf r},\sigma)\, d{\bf r}d\sigma,
\end{equation}
\begin{equation}
S_{\chi}[\rho]=-\int \rho({\bf r},\sigma)\ln\left\lbrack \frac{\rho({\bf r},\sigma)}{\chi(\sigma)}\right \rbrack\, d{\bf r}d\sigma,
\end{equation}
\begin{equation}
S[\overline{\omega}]=-\int C(\overline{\omega})\, d{\bf r},
\end{equation}
\begin{equation}
\Gamma_2^{c.g.}[\overline{\omega}]=\int \overline{\omega}^2\, d{\bf r},
\end{equation}
with $\chi(\sigma)\equiv {\rm exp}(-\sum_{n>1}\alpha_n\sigma^n)$ and $C(\overline{\omega})=-\int^{\overline{\omega}}\lbrack(\ln\hat{\chi})'\rbrack^{-1}(-x)\, dx$ where $\hat{\chi}(\Phi)\equiv \int\chi(\sigma)e^{-\sigma\Phi}\, d\sigma$.

In the framework of the MRS approach
where all the Casimirs are
conserved, the maximization of a ``generalized entropy''
$S[\overline{\omega}]$ at fixed energy and circulation provides a {\it
sufficient} condition of MRS thermodynamical stability
\cite{bouchet,proc,cnd}.  However, the reciprocal is wrong in case of
``ensemble inequivalence'' between microcanonical and grand
microcanonical ensembles. Indeed, the coarse-grained vorticity field $\overline{\omega}({\bf r})$
associated with a MRS thermodynamical equilibrium (i.e. a maximum of entropy $S[\rho]$ at fixed energy, circulation and Casimirs) is not necessarily a
maximum of generalized entropy $S[\overline{\omega}]$ at fixed energy and circulation (it can
be a saddle point of generalized entropy at fixed energy and
circulation).

In the framework of the EHT approach where the conservation
of the Casimirs is replaced by the specification of a prior vorticity
distribution (i.e. the Casimirs are treated canonically), 
the maximization of a generalized entropy $S[\overline{\omega}]$ at fixed
energy and circulation provides a {\it necessary and sufficient}
condition of EHT thermodynamical stability \cite{eht,aussois,cnd}. Indeed, a
vorticity distribution $\rho({\bf r},\sigma)$ is a EHT thermodynamical equilibrium (i.e. a maximum of relative entropy $S_{\chi}[\rho]$ at fixed energy and circulation) if and
only if the corresponding coarse-grained vorticity field $\overline{\omega}({\bf r})$ is a maximum
of generalized entropy $S[\overline{\omega}]$ at fixed energy and circulation.

Thus, we symbolically have
\begin{equation}
({\rm MRS})\Leftarrow ({\rm EHT})\Leftrightarrow ({\rm MaxS})
\label{conn1}
\end{equation}

Let us now specialize on the case of Gaussian distributions.

In the framework of the MRS approach where all the Casimirs are conserved, the minimization of macroscopic enstrophy $\Gamma_2^{c.g.}[\overline{\omega}]$ at fixed energy and circulation provides a {\it sufficient} condition of MRS thermodynamical stability for initial conditions leading to a Gaussian vorticity distribution at equilibrium \cite{bouchet,proc,cnd}. However, the reciprocal is wrong in case of
``ensemble inequivalence'', i.e. the coarse-grained vorticity field
$\overline{\omega}({\bf r})$ associated with a MRS thermodynamical
equilibrium with Gaussian vorticity distribution is not necessarily a minimum
enstrophy state (it can be a saddle point of macroscopic enstrophy at
fixed energy and circulation).

In the framework of the EHT approach where the conservation of the  Casimirs is replaced by the specification of a prior vorticity distribution, the minimization of macroscopic enstrophy $\Gamma_2^{c.g.}[\overline{\omega}]$ at fixed energy and circulation provides a {\it necessary and sufficient} condition of EHT thermodynamical stability for a Gaussian prior \cite{eht,aussois,cnd}, i.e. a
vorticity distribution $\rho({\bf r},\sigma)$ is a EHT thermodynamical equilibrium with a Gaussian prior if and
only if the corresponding coarse-grained vorticity field $\overline{\omega}({\bf r})$ is a minimum of macroscopic enstrophy $\Gamma_2^{c.g.}[\overline{\omega}]$ at fixed energy and circulation.

In the framework of the NCD approach where only the microscopic enstrophy $\Gamma_2^{f.g.}[\rho]$ is conserved among the Casimir constraints,
the minimization of macroscopic enstrophy
$\Gamma_2^{c.g.}[\overline{\omega}]$ at fixed energy and circulation
provides a {\it necessary and sufficient} condition of NCD
thermodynamical stability. Indeed, a vorticity distribution $\rho({\bf
r},\sigma)$ is a NCD thermodynamical equilibrium (i.e. a maximum of
entropy $S[\rho]$ at fixed energy, circulation and microscopic
enstrophy) if and only if the corresponding coarse-grained vorticity
field $\overline{\omega}({\bf r})$ is a minimum of macroscopic
enstrophy $\Gamma_2^{c.g.}[\overline{\omega}]$ at fixed energy and
circulation.

Thus, we symbolically have
\begin{equation}
({\rm MRS})\Leftarrow ({\rm EHT}) \Leftrightarrow  ({\rm Min}\Gamma_2)\Leftrightarrow ({\rm NCD})
\label{conn2}
\end{equation}

{\it Remark 1:} the EHT and NCD approaches provide {\it sufficient}
conditions of MRS stability. They are valuable in that
respect as they are simpler to solve. They may also have a deeper
physical meaning as discussed in the  introduction.

{\it Remark 2:} the equivalence between ({\rm EHT}) for a Gaussian prior  and ({\rm NCD}) is essentially coincidental because these variational problems are physically very different. In particular, this agreement is only valid for a Gaussian prior and does not extend to more general cases.

Finally, for completeness, we mention similar results obtained in \cite{k} for axisymmetric flows. In the framework of the Naso-Monchaux-Chavanis-Dubrulle (NMCD) approach where only the microscopic energy $E^{f.g.}[\rho]$ is conserved among the Casimir constraints,
the minimization of macroscopic energy $E^{c.g.}[\overline{\sigma}]$ at fixed helicity  and
angular momentum provides a {\it necessary and sufficient} condition of NMCD
thermodynamical stability, i.e. a distribution of angular momentum $\rho({\bf r},\eta)$ is a maximum of entropy $S[\rho]$ at fixed helicity, angular momentum and microscopic energy $E^{f.g.}[\rho]$ if and only if the corresponding
coarse-grained angular momentum distribution $\overline{\sigma}({\bf r})$ is a minimum of macroscopic energy $E^{c.g.}[\overline{\sigma}]$ at fixed
helicity and angular momentum.

Thus, we  symbolically have
\begin{equation}
({\rm NMCD}) \Leftrightarrow  ({\rm Min}E)
\label{conn3}
\end{equation}

\section{Phase transitions in 2D Euler flows}
\label{sec_ptg}

\subsection{Minimum enstrophy states}
\label{sec_mes}

In this section we study the maximization problem
\begin{equation}
\max_{\overline{\omega}}\lbrace S[\overline{\omega}]\, | \, E,\, \Gamma  \rbrace,\label{mes1}
\end{equation}
where $S=-\frac{1}{2}\int \omega^2\, d{\bf r}$ is the neg-enstrophy
(the opposite of the enstrophy), $E=\frac{1}{2}\int \omega\psi\,
d{\bf r}$ the energy and $\Gamma=\int
\omega\, d{\bf r}$ the circulation. The
maximization problem (\ref{mes1}) can be interpreted as: (i) a
criterion of nonlinear dynamical stability with respect to the 2D
Euler equation (Sec. \ref{sec_ss}), (ii) a phenomenological minimum
enstrophy principle (Sec. \ref{sec_ens}), (iii) a sufficient condition
of MRS thermodynamical stability \cite{bouchet,proc}, (iv) a necessary
and sufficient condition of EHT thermodynamical stability for a
Gaussian prior \cite{aussois,cnd}, (v) a necessary and sufficient
condition of thermodynamical stability in the energy-enstrophy-circulation
statistical theory where only the microscopic enstrophy is conserved
among the Casimirs (Sec. \ref{sec_equivgen}). For simplicity and
convenience, we shall call $S$ the entropy.

We write the variational principle for the first order variations as
\begin{eqnarray}
\label{mes2}
\delta S-\beta\delta E-\alpha\delta\Gamma=0,
\end{eqnarray}
where $\beta$ and $\alpha$ are Lagrange multipliers. This yields a linear $\omega-\psi$ relationship
\begin{eqnarray}
\label{mes3}
\omega=-\Delta\psi=-\beta\psi-\alpha.
\end{eqnarray}
As before, we assume that the area of
the domain is unity and we set $\langle X \rangle=\int X\, d{\bf
r}$. Taking the space average of Eq. (\ref{mes3}), we obtain  $\Gamma=-\beta\langle \psi\rangle-\alpha$ so that the
foregoing equation can be rewritten
\begin{eqnarray}
\label{mes4}
-\Delta\psi+\beta\psi=\Gamma+\beta\langle \psi\rangle,
\end{eqnarray}
with $\psi=0$ on the domain boundary\footnote{As
mentioned in the introduction, our approach assumes that forcing and
dissipation equilibrate each other so that the system becomes, in
average, statistically equivalent to the 2D Euler equation. Therefore,
we use boundary conditions consistent with the 2D Euler
equation. However, our approach is expected to be valid only in the
bulk of the flow, relatively far from the boundary layers where our
assumptions do not hold anymore. Therefore, the boundary that we
consider may not correspond to the true, physical, boundary of the
fluid, but may be an ``effective domain'' where our inviscid assumption
applies. For further discussion on the influence of boundary conditions
on the structure of the flow, see Keetels {\it et al.} \cite{keetels}
and references therein.}. This is the fundamental differential
equation of the problem. The energy and the entropy can then be
expressed as
\begin{eqnarray}
\label{mes5}
E=-\frac{1}{2}\beta\left (\langle \psi^2\rangle-\langle \psi\rangle^2\right )+\frac{1}{2}\Gamma\langle \psi\rangle,
\end{eqnarray}
\begin{eqnarray}
\label{mes6}
S=-\frac{1}{2}\beta^2\left (\langle \psi^2\rangle-\langle \psi\rangle^2\right )-\frac{1}{2}\Gamma^2.
\end{eqnarray}
We shall study the maximization problem (\ref{mes1}) by adapting the approach
of Chavanis \& Sommeria \cite{jfm} to this specific situation (these authors
studied a related but not exactly equivalent problem).  We will see that the structure of the problem depends on a unique control parameter \cite{jfm}:
\begin{eqnarray}
\label{mes7}
\Lambda=\frac{\Gamma}{\sqrt{2E}}.
\end{eqnarray}
We note that the maximization problem (\ref{mes1}) has been studied recently  by Venaille \&
Bouchet \cite{vb} by using a different theoretical treatment. They performed a detailed analysis of the phase transitions associated with (\ref{mes1}) in the context of statistical mechanics, emphasizing in particular the notion of ensemble inequivalence. However, their approach is very abstract.  Our study is more direct and  can offer a complementary discussion of the problem. The maximization problem (\ref{mes1}) has also been studied recently by Keetels {\it et al.} \cite{keetels} with different boundary conditions adapted to viscous flows.

\subsection{The bifurcation diagram} \label{sec_ds}

In this section, we apply the methodology developed by Chavanis \& Sommeria \cite{jfm}. This methodology is relatively general: it is valid for an arbitrary domain and for an arbitrary linear operator. However, for illustration, we shall consider the Laplacian operator and a rectangular domain.

\subsubsection{The eigenmodes} \label{sec_e}

We first assume that
\begin{eqnarray}
\label{e1}
\Gamma+\beta\langle \psi\rangle=0,
\end{eqnarray}
corresponding to $\alpha=0$. In that case, the differential equation (\ref{mes4}) becomes
\begin{eqnarray}
\label{e2}
-\Delta\psi+\beta\psi=0,
\end{eqnarray}
with $\psi=0$ on the domain boundary. Using the results of Appendix \ref{sec_eigen}, Eq. (\ref{e2}) has solutions only for $\beta=\beta_{mn}$ (eigenvalues) and the corresponding solutions (eigenfunctions) are
\begin{eqnarray}
\label{e4}
\psi=\left (\frac{2E}{-\beta_{mn}}\right )^{1/2}\psi_{mn},
\end{eqnarray}
where we have used the energy constraint (\ref{mes5}) to determine the normalization constant. Substituting this result in Eq. (\ref{e1}), we find that these solutions exist only for $\Lambda=\Lambda_{mn}$ with
\begin{eqnarray}
\label{e6}
\Lambda_{mn}^2=-\beta_{mn}\langle\psi_{mn}\rangle^2.
\end{eqnarray}
For the eigenmodes $\langle \psi_{mn}\rangle=0$ ($m$ or $n$ even), we find $\Lambda=0$ and for the  eigenmodes $\langle \psi_{mn}\rangle\neq 0$ ($m$ and $n$ odd), we find $\Lambda^2={\Lambda''_{mn}}^2\equiv -\beta_{mn}\langle\psi_{mn}\rangle^2\neq 0$.

\subsubsection{The solutions of the continuum} \label{sec_c}

We now assume that $\Gamma+\beta\langle \psi\rangle\neq 0$ and we define
\begin{eqnarray}
\label{c1}
\phi=\frac{\psi}{\Gamma+\beta\langle \psi\rangle}.
\end{eqnarray}
In that case, the differential equation (\ref{mes4}) becomes
\begin{eqnarray}
\label{c2}
-\Delta\phi+\beta\phi=1,
\end{eqnarray}
with $\phi=0$ on the domain boundary.  We also assume that $\beta\neq \beta_{mn}$. In that case, Eq. (\ref{c2}) has a unique solution that can be obtained by expanding $\phi$ on the eigenmodes. We get
\begin{eqnarray}
\label{c3}
\phi=\sum_{mn} \frac{\langle\psi_{mn}\rangle}{\beta-\beta_{mn}}\psi_{mn},
\end{eqnarray}
where only the modes with $\langle\psi_{mn}\rangle\neq 0$ are ``excited''.

For $\Gamma\neq 0$, taking the average of Eq. (\ref{c1}) and solving for $\langle\psi\rangle$, we obtain
$\langle \psi\rangle=\Gamma\langle \phi\rangle/(1-\beta \langle \phi\rangle)$. Therefore, the solution of Eq. (\ref{mes4}) is
\begin{eqnarray}
\label{c4}
\psi=\frac{\Gamma\phi}{1-\beta \langle \phi\rangle}.
\end{eqnarray}
Substituting this solution in the energy constraint (\ref{mes5}), we obtain the ``equation of state'':
\begin{eqnarray}
\label{c5}
(1-\beta\langle\phi\rangle)^2=\Lambda^2 (\langle\phi\rangle-\beta\langle\phi^2\rangle).
\end{eqnarray}
This equation determines $\beta$ as a function of $\Lambda$. In particular, it determines the caloric curve $\beta(E)$ for a given value of $\Gamma\neq 0$. Note that the equation of state involves the important function \cite{jfm}:
\begin{eqnarray}
\label{imp}
F(\beta)\equiv \beta\langle\phi\rangle-1.
\end{eqnarray}

For $\Gamma=0$, the solution of Eq. (\ref{mes4}) is
\begin{eqnarray}
\label{c7}
\psi=\beta\langle\psi\rangle\phi.
\end{eqnarray}
Taking the space average of this relation, we find that this solution exists only for a discrete set of temperatures $\beta=\beta_*^{(k)}$ satisfying $F(\beta_*^{(k)})=0$.
We shall note $\beta_*\equiv\beta_*^{(1)}$ the largest of these solutions. Substituting Eq. (\ref{c7}) in the energy constraint (\ref{mes5}), we find that the amplitude $\langle\psi\rangle$ is determined by
\begin{eqnarray}
\label{c9}
E=-\frac{1}{2}\beta^3\langle\psi\rangle^2 (\langle\phi^2\rangle-\langle\phi\rangle^2).
\end{eqnarray}
Of course, the case $\Lambda=0$ is also a limit case of the equation of state (\ref{c5}).

\subsubsection{The mixed solutions} \label{sec_m}

For $\beta\rightarrow \beta_{mn}$ with $m$ and $n$ odd ($\langle\psi_{mn}\rangle\neq 0$), we find from Eq. (\ref{c3}) that $\phi\sim \langle\psi_{mn}\rangle\psi_{mn}/(\beta-\beta_{mn})\rightarrow +\infty$ leading to $\Lambda\rightarrow \Lambda''_{mn}$ and $\psi\rightarrow  (2E/\beta_{mn})^{1/2} \psi_{mn}$. Therefore, we recover the eigenfunction $\psi_{mn}$ as a limit case.
The eigenfunctions with non vanishing average value are therefore contained on the continuum branch.

For  $\beta=\beta_{mn}$ with $m$ or $n$ even ($\langle\psi_{mn}\rangle= 0$), the solution of Eq. (\ref{c2}) is not unique. It corresponds to the mixed solutions
\begin{eqnarray}
\label{m1}
\phi=\sum_{m'n'} \frac{\langle\psi_{m'n'}\rangle}{\beta_{mn}-\beta_{m'n'}}\psi_{m'n'}+\chi_{mn}\psi_{mn},
\end{eqnarray}
where $\chi_{mn}$ is determined by the energy constraint (more precisely, it can be related to $\Lambda$ by substituting Eq. (\ref{m1}) in Eq. (\ref{c5}) where now $\beta=\beta_{mn}$). These solutions form a plateau at fixed temperature $\beta=\beta_{mn}$. For  $\chi_{mn}\rightarrow +\infty$, we recover the pure eigenmode $\psi_{mn}$ that exists at $\Lambda=0$ and for  $\chi_{mn}=0$, we connect the branch of continuum solutions at $\Lambda=\Lambda'_{mn}$.

The general bifurcation diagram showing the eigenmodes, the solutions of the continuum and the mixed solutions is represented in Fig. 2 of \cite{jfm} (see also Figs. \ref{fig_2} and \ref{fig_3} below).

\subsection{The geometry induced monopole/dipole transition} \label{sec_pt}

For a given value of the control parameter $\Lambda$, we can have an infinite number of solutions to Eq. (\ref{mes4}) \cite{jfm}. We can now use the  entropy (\ref{mes6}) to select the most probable state (maximum entropy state) among all these solutions.

For the eigenmodes, the entropy takes the simple form $S/E=\beta_{mn}$. In particular, for the eigenmodes $\psi_{mn}$ with $m$ or $n$ even ($\langle\psi_{mn}\rangle=0$) that exist only for $\Lambda=0$, we have
\begin{eqnarray}
\label{pt1}
\frac{S}{E}(\Lambda=0)=\beta_{mn}.
\end{eqnarray}
For a rectangular domain elongated in the $x$ direction, the eigenmode with the highest entropy at $\Gamma=0$ is the dipole $(m,n)=(2,1)$ with temperature $\beta_{21}(\tau)$. Therefore, the maximum entropy state (or the minimum enstrophy state) corresponds to the mode with the largest scale. The modes with smaller scales ($m,n$ large) have lower entropy (higher enstrophy). Therefore, the maximum entropy and minimum enstrophy principles select the large-scale structures among the infinite class of steady states of the 2D Euler equation. This is a manifestation of the inverse cascade process.

For the solutions of the continuum, the entropy can be written
\begin{eqnarray}
\label{pt2}
S/E=\beta \left (1+\Lambda^2 \frac{\langle\phi\rangle}{\beta\langle\phi\rangle-1}\right )-\Lambda^2.
\end{eqnarray}
For $\Lambda=0$, this expression reduces to
\begin{eqnarray}
\label{pt3}
\frac{S}{E}(\Lambda=0)=\beta_{*}^{(k)}.
\end{eqnarray}
The solution with highest entropy is the monopole with temperature $\beta_{*}(\tau)$.

\begin{figure}[h]
\center
\includegraphics[width=8cm,keepaspectratio]{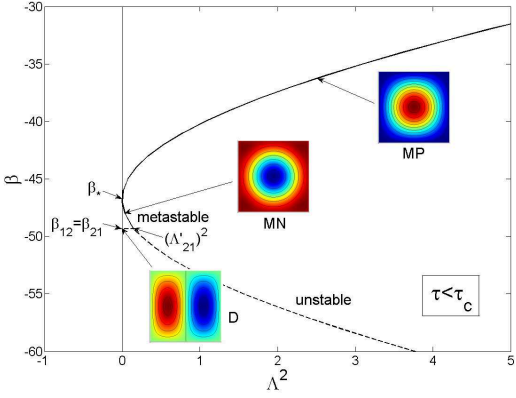}
\caption{\label{fig_2} Series of equilibria  in a square domain ($\tau=1<\tau_c$). In that case $\max\lbrace{\beta_*,\beta_{21}\rbrace}=\beta_*$. The maximum entropy state is the direct monopole (for $\Gamma>0$ the vorticity is positive at the center and negative at the periphery (MP); for $\Gamma<0$ this is the opposite (MN)) for any value of $\Lambda^2$. For $\Lambda^2<(\Lambda'_{21})^2$, the reversed monopole is metastable (local entropy maximum) as discussed in Sec. \ref{sec_sa}. Note that the metastable states have negative specific heats $C=\frac{\partial E}{\partial T}=\beta^2E^2\frac{\partial(1/E)}{\partial\beta}<0$. For $\Gamma=0$, the direct and reversed monopoles have the same entropy. For fixed $\Gamma$, the caloric curve $\beta(E)$ does not present any phase transition (see Sec. \ref{sec_dpt}). The vorticity profiles are plotted for $\Gamma\ge 0$. The red colour corresponds to positive values of the vorticity and the blue colour to negative values.}
\end{figure}

\begin{figure}[h]
\center
\includegraphics[width=8cm,keepaspectratio,angle=0]{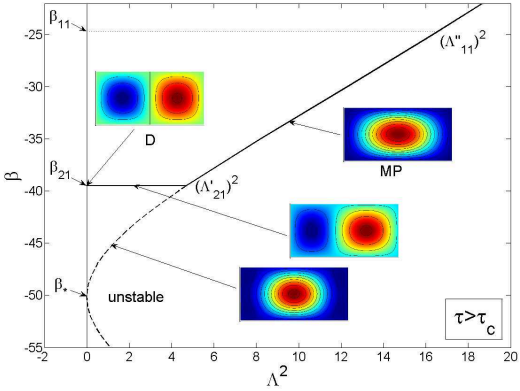}
\caption{\label{fig_3} Series of equilibria  in a rectangular  domain with aspect ratio $\tau=2>\tau_c$. In that case  $\max\lbrace{\beta_*,\beta_{21}\rbrace}=\beta_{21}$. The maximum entropy state is the dipole for  $\Lambda^2<(\Lambda'_{21})^2$ and the direct monopole for $\Lambda^2>(\Lambda'_{21})^2$ (the reversed monopoles are unstable). For $\Gamma\neq 0$, the caloric curve $\beta(E)$ presents a second order phase transition marked by the discontinuity of $\frac{\partial\beta}{\partial E}(E)$ at $E=E'_{21}$ as discussed in Sec. \ref{sec_dpt}.}
\end{figure}

For $\Lambda=0$, we have a competition between the monopole $\beta_*(\tau)$ (continuum branch) and the dipole $\beta_{21}(\tau)$ (eigenmode)\footnote{{As noted by Taylor {\it et al.} \cite{taylor}, the solution with $\langle\psi\rangle=0$ (dipole) has zero angular momentum while the solution with $\langle\psi\rangle\neq 0$ (monopole) has nonzero angular momentum, even though the circulation is zero. This explains the ``spin-up'' phenomenon discovered in \cite{spin}.}}.  We must therefore compare their entropy (or equivalently their inverse temperature) to select the maximum entropy state.  As shown in Chavanis \& Sommeria \cite{jfm}, this selection depends on the geometry of the domain. In a rectangular domain, it is found that the monopole has the highest entropy ($\beta_*>\beta_{21}$) for $\tau<\tau_c=1.12$ while the dipole dominates ($\beta_{21}>\beta_{*}$) for $\tau>\tau_c=1.12$. More generally, it can be shown that the entropy is a monotonically increasing function of the inverse temperature (for a fixed value of $\Lambda$). Therefore, at any $\Lambda$, the maximum entropy state is the one with the highest inverse temperature \cite{jfm}.  The series of equilibria $\beta(\Lambda)$  is represented in Figs.~\ref{fig_2} and \ref{fig_3} for a square domain and for a rectangular domain of aspect ratio $2$, respectively. For $\tau<\tau_c$, the maximum entropy state is the direct monopole for any value of $\Lambda$. For $\tau>\tau_c$, the maximum entropy state is the dipole for $\Lambda^2<(\Lambda'_{21})^2$ and the direct monopole for $\Lambda^2>(\Lambda'_{21})^2$.

\subsection{Stability analysis and ensemble inequivalence}
\label{sec_sa}

For given values of $E$ and $\Gamma$, there can exist different
critical points of entropy $S$ (canceling its first order
variations). They are solutions of the differential equation
(\ref{mes4}). For sufficiently small $\Lambda$, there exists an
infinity of solutions \cite{jfm}. In the last section, we have
compared the value of the entropy of these different solutions in
order to select the maximum entropy state. However, a more precise
study should determine which solutions correspond to global entropy
maxima, local entropy maxima and saddle points. Saddle points of
entropy are unstable and should be rejected in principle (see, however
Sec. \ref{sec_per}). By contrast, local entropy maxima (metastable
states) can be long-lived for systems with long-range interactions. In
practice, they are as much relevant as global entropy maxima (stable
states). In the following, using an approach very close to the one
followed by Chavanis \& Sommeria \cite{jfm} (but not exactly
equivalent since the variational problems differ), we determine
sufficient conditions of {\it instability}. This will eliminate a
large class of solutions that are unstable saddle points of entropy
and give the form of the perturbations that destabilize them. The
remaining solutions are either stable or metastable.

A critical point of entropy at fixed energy and circulation is a (local) maximum iff
\begin{eqnarray}
\label{sa1}
\delta^2 J=-\int \frac{(\delta\omega)^2}{2}\, d{\bf r}-\frac{1}{2}\beta\int\delta\omega\delta\psi\, d{\bf r}<0,
\end{eqnarray}
for all perturbations that conserve energy and circulation at first order: $\delta E=\delta \Gamma=0$.

(i) We first show that all the solutions with $\beta<\beta_{21}$ are unstable (saddle points). To that purpose, we consider a perturbation of the form $\delta\omega=\psi_{21}({\bf r})$. The corresponding stream function is $\delta\psi=-\frac{1}{\beta_{21}}\psi_{21}({\bf r})$. For this perturbation, it is clear that $\delta\Gamma=\int \delta\omega\, d{\bf r}=0$ since $\langle\psi_{21}\rangle=0$. Furthermore, $\delta E=\int \psi\delta\omega\, d{\bf r}=0$ since $\psi_{21}$ is orthogonal to the other eigenmodes and to the solutions of the continuum (as they involve a summation (\ref{c3}) on the eigenmodes with non zero average  that are orthogonal to $\psi_{21}$). Finally, a simple calculation shows that
\begin{eqnarray}
\label{sa2}
\delta^2 J=\frac{1}{2}\left (\frac{\beta}{\beta_{21}}-1\right )>0.
\end{eqnarray}
We have thus found a particular perturbation that increases the entropy at fixed energy and circulation. Therefore, the states with $\beta<\beta_{21}$ are unstable (saddle points).

(ii) We now show that if $\beta_{21}<\beta_*$ the mode $\psi_{21}$ existing at $\Lambda=0$ is unstable (saddle point). To that purpose, we consider a perturbation of the form $\delta\omega=1-\beta_* \phi_*$ (where $\phi_*$ is the solution of Eq. (\ref{c2}) corresponding to $\beta=\beta_*$).  The corresponding stream function is $\delta\psi=\phi_*$. For this perturbation, it is clear that $\delta\Gamma=\int \delta\omega\, d{\bf r}=0$ since $1-\beta_*\langle\phi_*\rangle=0$. Furthermore, $\delta E=\int \psi_{21}\delta\omega\, d{\bf r}=0$ since $\langle\psi_{21}\rangle=0$ and $\psi_{21}$ is perpendicular to $\phi_*$ as explained previously. Finally, after some simple algebra using $1-\beta_*\langle\phi_*\rangle=0$, we get
\begin{eqnarray}
\label{sa3}
\delta^2 J=\frac{1}{2}(\beta_*-\beta_{21})(\langle\phi_*\rangle-\beta_* \langle\phi_*^2\rangle)>0,
\end{eqnarray}
(the last term in parenthesis is positive as shown in Appendix \ref{sec_eigen}). We have thus found a particular perturbation that increases the entropy at fixed energy and circulation. Therefore, if $\beta_{21}<\beta_*$ the mode $\psi_{21}$ existing at $\Lambda=0$ is unstable. By continuity, the mixed solutions forming a plateau at $\beta=\beta_{21}$ are also
unstable if $\beta_{21}<\beta_*$ since the two ends of the plateau are unstable.

The maximization problem (\ref{mes1}) corresponds to a condition of
microcanonical stability which is relevant to our problem since the
circulation and the energy are conserved by the 2D Euler
equation. However, it can be convenient to establish criteria of
canonical and grand canonical stability. Indeed, the solution of a
maximization problem is always solution of a more constrained dual
maximization problem, but the reciprocal is wrong in case of ensemble
inequivalence that is generic for systems with long-range interactions
\cite{ellis}.  Therefore, conditions of canonical and grand canonical
stability provide only {\it sufficient} conditions of microcanonical
stability: grand canonical stability implies canonical stability which
itself implies microcanonical stability (see, e.g., \cite{proc}).
This problem of ensemble inequivalence has been studied in detail by
Venaille \& Bouchet \cite{vb} and we briefly discuss it again bringing
some complements regarding the metastable states (that are not
considered in \cite{vb}).

Considering the grand canonical ensemble, we have to maximize the grand potential $G=S-\beta E-\alpha\Gamma$ (no constraint problem). The condition of grand canonical stability corresponds to inequality (\ref{sa1}) for all variations $\delta\omega$. By decomposing the perturbation of the eigenmodes of the Laplacian, it is easy to show that the system is a maximum of grand potential iff $\beta>\beta_{11}$ (where $\beta_{11}$ is the largest eigenvalue of the Laplacian). This is closely related to the Arnol'd theorem (indeed, the grand potential is equivalent to the Arnol'd energy-Casimir functional \cite{proc}; furthermore, for a linear $\omega-\psi$ relationship, the Arnol'd theorem, which usually provides only a sufficient condition of grand canonical stability, now provides a necessary and sufficient condition of grand canonical stability). Since grand canonical stability implies microcanonical stability (but not the converse) we conclude that, if $\beta>\beta_{11}$, the system is a maximum of entropy at fixed circulation and energy.

Considering the canonical ensemble, we have to maximize the free
energy $J=S-\beta E$ at fixed circulation (one constraint
problem). The condition of canonical stability corresponds to
inequality (\ref{sa1}) for all variations $\delta\omega$ that conserve
circulation. By carefully taking into account the constraint on the
circulation, Venaille \& Bouchet \cite{vb} show that the system is a
maximum of free energy iff $\beta>\max\lbrace
\beta_{21},\beta_*\rbrace$. In particular, the states with
$\max\lbrace \beta_{21},\beta_*\rbrace<\beta<\beta_{11}$ are stable in
the canonical ensemble but unstable in the grand canonical
ensemble. Thus, canonical and grand canonical ensembles are
inequivalent \cite{vb}. On the other hand, since canonical stability
implies microcanonical stability (but not the converse) we conclude
that, if $\beta>\max\lbrace \beta_{21},\beta_*\rbrace$, the system is
a maximum of entropy at fixed circulation and energy. In particular,
the states with $E>\Gamma^2/(2\Lambda_{11}^2)\equiv E_{11}(\Gamma)$
are stable in the canonical ensemble but unstable in the grand
canonical ensemble \cite{vb}. Note that the states with
$\beta<\beta_*$ are unstable in the canonical ensemble (they are
saddle points of free energy at fixed circulation). This result can be
obtained directly by considering a perturbation of the form
$\delta\omega=1-\beta_* \phi_*$ like in (ii). For this perturbation,
$\delta\Gamma=0$. On the other hand, in the canonical ensemble, we do
not need to impose $\delta E=0$ so that this perturbation can be
applied to {\it any} state leading to
\begin{eqnarray}
\label{sa3bis}
\delta^2 J=\frac{1}{2}(\beta_*-\beta)(\langle\phi_*\rangle-\beta_* \langle\phi_*^2\rangle)>0,
\end{eqnarray}
which proves the result. By contrast, this argument does not work in the microcanonical ensemble since
the chosen perturbation does not satisfy $\delta E=0$ for all states. Therefore, when $\beta_*>\beta_{21}$, the states with $\beta_{21}<\beta<\beta_*$ are unstable in the canonical ensemble (they are saddle points of free energy at fixed circulation) while they are  metastable in the microcanonical ensemble (they are local maxima of entropy at fixed circulation and energy). This is an interesting notion of ensemble inequivalence which affects metastable states (Venaille \& Bouchet \cite{vb} show that the microcanonical and canonical ensembles are equivalent for the fully stable states but the case of metastable states is not considered in their study). In particular, we note that the metastable states with $\beta_{21}<\beta<\beta_*$ have {\it negative specific heats} (see Fig. \ref{fig_2}). This is allowed  in the microcanonical ensemble but not in the canonical ensemble. Interestingly, this is the first observation of negative specific heats in that context.

Combining all these results, we conclude that in the microcanonical ensemble:

(a) If $\beta_{21}<\beta_*$: the states are stable for $\beta\ge \beta_*$, unstable for $\beta\le \beta_{21}$ and metastable for  $\beta_{21}<\beta<\beta_*$, as shown in Fig.~\ref{fig_2}. Therefore, the direct monopole  is stable for any $\Lambda^2$  and the reversed monopole is metastable for $\Lambda^2<(\Lambda'_{21})^2$.

(b) If $\beta_{21}>\beta_*$: the states are stable for $\beta\ge \beta_{21}$ and unstable for $\beta< \beta_{21}$, as shown in Fig.~\ref{fig_3}. Therefore, the dipole is stable for  $\Lambda^2<(\Lambda'_{21})^2$ and the direct monopole is stable for $\Lambda^2>(\Lambda'_{21})^2$. There is no metastable state in that case.

\subsection{The chemical potential} \label{sec_cp}

\begin{figure}[h]
\center
\includegraphics[width=8cm,keepaspectratio]{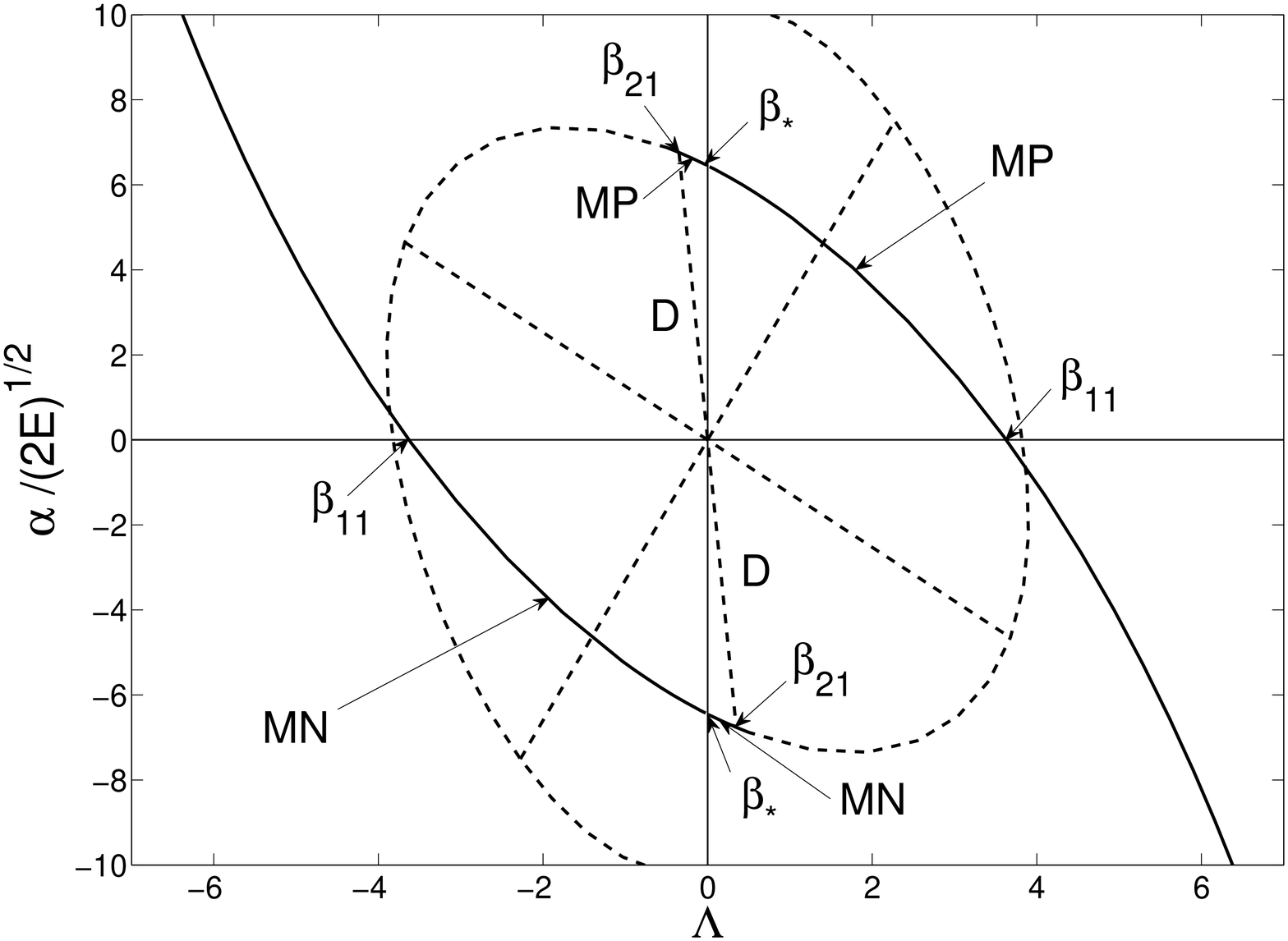}
\caption{\label{fig_4} Relationship between $\alpha/\sqrt{2E}$ and $\Lambda$ in a square domain (case $\beta_{21}<\beta_*$). The solid lines correspond to the stable ($\beta\ge\beta_*$) and metastable ($\beta_{21}<\beta<\beta_*$) states. Unstable states ($\beta\le\beta_{21}$) are represented by the dashed lines. The straight lines represent the mixed solutions with constant temperature: $\beta=\beta_{12}=\beta_{21}$, $\beta=\beta_{22}$, $\beta=\beta_{23}$.}
\end{figure}

\begin{figure}[h]
\center
\includegraphics[width=8cm,keepaspectratio]{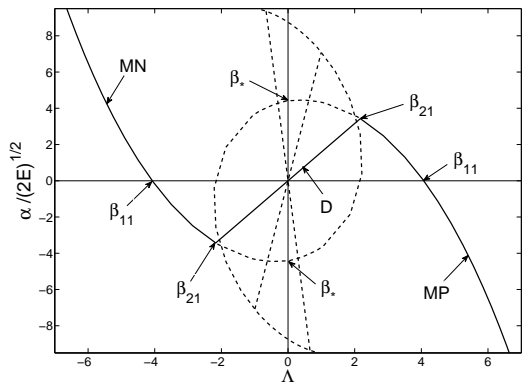}
\caption{\label{fig_5} Relationship between $\alpha/\sqrt{2E}$ and $\Lambda$ in a rectangular domain of aspect ratio 2 (case $\beta_{21}>\beta_*$). The solid lines correspond to the stable states ($\beta\ge\beta_{21}$). Unstable  states ($\beta<\beta_{21}$) are represented by the dashed lines. The straight lines represent the mixed solutions with constant temperature: $\beta=\beta_{21}$, $\beta=\beta_{12}$, $\beta=\beta_{22}$.}
\end{figure}

In Sec. \ref{sec_pt}, we have represented the inverse temperature $\beta$ as a function of $\Lambda$. We shall now study how the chemical potential $\alpha$ depends on $\Lambda$. The chemical potential is given by $\alpha=-\beta\langle\psi\rangle-\Gamma$.
For the eigenmodes,
\begin{eqnarray}
\label{cp2}
\alpha=0.
\end{eqnarray}
For the solutions of the continuum, assuming $\Gamma\neq 0$, and using Eq. (\ref{c4}), we get $\alpha={\Gamma}/(\beta\langle\phi\rangle-1)={\Gamma}/F(\beta)$.
Therefore,
\begin{eqnarray}
\label{cp4}
\frac{\alpha}{\sqrt{2E}}=\frac{\Lambda}{F(\beta)}.
\end{eqnarray}
For $\Gamma=0$, using Eq. (\ref{c7}), we obtain
\begin{eqnarray}
\label{cp5}
\frac{\alpha}{\sqrt{2E}}=\pm\frac{1}{\sqrt{-\beta_* (\langle\phi_*^2\rangle-\langle\phi_*\rangle^2)}},
\end{eqnarray}
which is a limit case of Eq. (\ref{cp4}). The normalized chemical potential ${\alpha}/{\sqrt{2E}}$ is plotted as a function of $\Lambda$ in Figs.~\ref{fig_4} and \ref{fig_5}, for a square domain and for a rectangular domain of aspect ratio $2$, respectively. To plot this curve, we have used Eqs. (\ref{c5}) and (\ref{cp4}).
For a given value of $\beta$, we can determine $\Lambda$ by Eq. (\ref{c5}) and ${\alpha}/{\sqrt{2E}}$ by  Eq. (\ref{cp4}). Therefore, we can obtain ${\alpha}/{\sqrt{2E}}$ as a function of $\Lambda$ parameterized by $\beta$ for the solutions of the continuum. For the mixed solutions, $\beta=\beta_{mn}$ is fixed and ${\alpha}/{\sqrt{2E}}$ is a linear function of $\Lambda$ given by Eq. (\ref{cp4}).

In Figs.~\ref{fig_4} and \ref{fig_5}, we have represented the series of equilibria containing all the critical points of entropy. If we continue the series of equilibria to more and more unstable states, the curve rolls up several times around the origin (not shown). As indicated above, the series of equilibria is parameterized by $\beta$. The branches corresponding to $\beta>\beta_{11}$ are stable in the grand canonical, canonical and microcanonical ensembles and the branches corresponding to $\beta>\max\lbrace \beta_{21},\beta_*\rbrace$ are stable in the canonical and microcanonical ensembles. The part of the branches corresponding to $\max\lbrace \beta_{21},\beta_*\rbrace<\beta<\beta_{11}$ are stable in the canonical and microcanonical ensembles but not in the grand canonical ensemble. For $\tau<\tau_c$ ($\beta_{21}<\beta_*$), the part of the branches corresponding to $\beta_{21}<\beta<\beta_*$ are metastable in the microcanonical ensemble and unstable in the other ensembles.

{\it Remark:} in the grand canonical ensemble, the control parameter
is the chemical potential $\alpha$ and the conjugated variable is the
circulation $\Gamma$. We must therefore consider $\Gamma(\alpha)$ by
rotating the curves of Figs. \ref{fig_4} and
\ref{fig_5} by $90^o$. Only the part of the curve with
$\beta>\beta_{11}$ (NW and SE quadrants) are stable in the grand
canonical ensemble. There is a first order grand canonical phase
transition at $\alpha=0$ marked by the discontinuity of the
circulation $\Gamma(\alpha)$ between
$\Gamma=\pm\Lambda''_{11}\sqrt{2E}$. Note that there is no metastable
states in the grand canonical ensemble because the states with
$\beta<\beta_{11}$ are all unstable.

\subsection{Description of phase transitions}
\label{sec_dpt}

We briefly discuss the nature of phase transitions associated with the maximization problem (\ref{mes1}) and confirm the results that Venaille \& Bouchet \cite{vb} obtained by a different method. We also give a special attention to the metastable states that are not considered in \cite{vb}.

We shall describe successively the caloric curve $\beta(E)$ for a fixed $\Gamma$ and the chemical potential curve $\alpha(\Gamma)$ for a fixed $E$. As first observed by Chavanis \& Sommeria \cite{jfm}, the nature of the phase transitions depends on the value of $\max\lbrace{\beta_*,\beta_{21}\rbrace}$. In a rectangular domain, this quantity is determined by  the value of the aspect ratio $\tau$. We must therefore consider two cases successively: $\tau<\tau_c$ and $\tau>\tau_c$.

\subsubsection{Caloric curve}

The caloric curve corresponds to the stable part of the series of equilibria $\beta(E)$ containing global (stable) and local (metastable) maximum entropy states at fixed $E$ and $\Gamma$.

$\bullet$ Let us first consider $\tau<\tau_c$ corresponding to
$\max\lbrace{\beta_*,\beta_{21}\rbrace}=\beta_*$ as in
Fig. \ref{fig_2}. For $\Gamma=0$, the maximum entropy state is the
monopole and the caloric curve is simply a straight line
$\beta(E,\Gamma=0)=\beta_*$. For each value of the energy, we have two
solutions with the same inverse temperature $\beta_*$ but different
values of the chemical potential $\alpha(\Gamma=0,E)=\pm\alpha_0$
(see Fig. \ref{fig_4}). One solution is a monopole
with positive vorticity at the center (MP) and the other solution is a
monopole with negative vorticity at the center
(MN). For $\Gamma=0$, these solutions have the same
entropy. Thus, the branch $\beta(E,\Gamma=0)=\beta_*$ is degenerate.
For $\Gamma\neq 0$, the caloric curve $\beta(E,\Gamma\neq 0)$ can be
deduced easily from Fig. \ref{fig_2}\footnote{In fact, it is more
convenient to plot $\beta$ as a function of the inverse of the energy
$1/E$ as in Fig. \ref{fig_2} since the interesting bifurcations occurs
for large values of the energy.}. The global maximum entropy state is
the direct monopole (for $\Gamma>0$ the vorticity is positive at the
center (MP); for $\Gamma<0$ the vorticity is negative at the center
(MN)) for any $E$. For $E>E'_{21}(\Gamma)\equiv
\frac{\Gamma^2}{2(\Lambda'_{21})^2}$, the reversed monopole is
metastable (local entropy maximum). Note that the metastable states
have negative specific heats $C\equiv dE/d(1/\beta)<0$. The caloric
curve $\beta(E)$ does not present any phase transition.

$\bullet$ Let us now consider $\tau>\tau_c$ corresponding to
$\max\lbrace{\beta_*,\beta_{21}\rbrace}=\beta_{21}$ as in
Fig. \ref{fig_3}. For $\Gamma=0$, the maximum entropy state is the
dipole and the caloric curve is simply a straight line
$\beta(E,\Gamma=0)=\beta_{21}$. For each value of the energy, we have
two solutions with the same inverse temperature and the same chemical
potential $\alpha(\Gamma=0,E)=0$ (see
Fig. \ref{fig_5}). One solution is a dipole $(+,-)$ with positive
vorticity on the left and the other solution is a dipole $(-,+)$ with
negative vorticity on the left (in Fig. \ref{fig_3}, we have only
represented the dipole $(-,+)$). For $\Gamma=0$, these
solutions have the same entropy. Thus, the branch
$\beta(E,\Gamma=0)=\beta_{21}$ is degenerate. For $\Gamma\neq 0$, the
caloric curve $\beta(E,\Gamma\neq 0)$ can be deduced easily from
Fig. \ref{fig_3}. The maximum entropy state is the
asymmetric (mixed) dipole $(+,-)$ or $(-,+)$ for
$E>E'_{21}(\Gamma)\equiv \frac{\Gamma^2}{2(\Lambda'_{21})^2}$ and the
direct monopole for $E<E'_{21}(\Gamma)\equiv
\frac{\Gamma^2}{2(\Lambda'_{21})^2}$ (the reversed monopoles are
unstable). The caloric curve $\beta(E)$ presents a second order phase
transition marked by the discontinuity of
$\frac{\partial\beta}{\partial E}(E)$ at $E=E'_{21}(\Gamma)$.

\subsubsection{Chemical potential curve}

The chemical potential curve corresponds to the stable part of the
series of equilibria $\alpha(\Gamma)$ containing global (stable) and
local (metastable) maximum entropy states at fixed $E$ and $\Gamma$.

$\bullet$ Let us first consider $\tau<\tau_c$ corresponding to
$\max\lbrace{\beta_*,\beta_{21}\rbrace}=\beta_*$ as in
Fig. \ref{fig_4}. The global maximum entropy state is the monopole for
any value of $\Gamma$. Considering only fully stable states (global
entropy maxima), there is a first order phase transition at $\Gamma=0$
marked by the discontinuity of $\alpha(\Gamma)$ while the entropy is
continuous.  When we pass from positive $\Gamma$ to negative $\Gamma$,
we pass discontinuously (in terms of $\alpha$ but not in terms of
$\beta$ or $S$) from the monopole (MP) to the monopole (MN). In fact,
due to the presence of long-lived metastable states (see
Sec. \ref{sec_hyst}), we remain in practice on the monopole (MP) until
the metastable branch disappears. Then we jump on the monopole (MN)
with discontinuity of $\alpha$ (and $\beta$ and $S$). This corresponds
to a zeroth order phase transition.

$\bullet$ Let us now consider $\tau>\tau_c$ corresponding to
$\max\lbrace{\beta_*,\beta_{21}\rbrace}=\beta_{21}$ as in
Fig. \ref{fig_5}. The global maximum entropy state is the
asymmetric (mixed) dipole for
$|\Gamma|<\Gamma'_{21}(E)\equiv \sqrt{2E}\Lambda'_{21}$ and the direct
monopole for $|\Gamma|>\Gamma'_{21}(E)$. There are two second order
phase transitions marked by the discontinuity of
$\frac{\partial\alpha}{\partial \Gamma}(\Gamma)$ at
$\Gamma=\pm\Gamma'_{21}(E)$.

\subsubsection{Phase diagram}

\begin{figure}[h]
\center
\includegraphics[width=8cm,keepaspectratio]{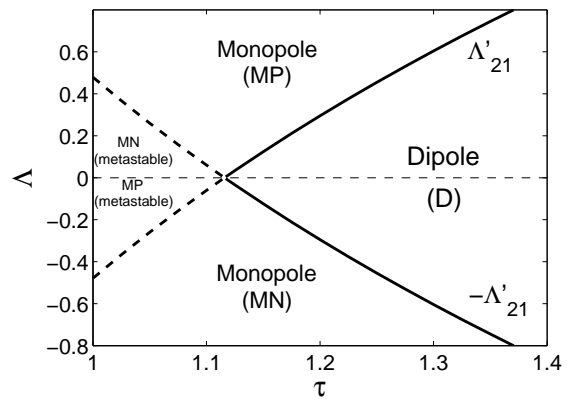}
\caption{\label{fig_8} Phase diagram in the $(\tau,\Lambda)$ plane showing the domain of stability of the direct monopoles and dipoles.  We have indicated by a dashed line the domain of metastability of the reversed monopoles.}
\end{figure}

The phase diagram in the $(\tau,\Lambda)$ plane, including the metastable states, is plotted in Fig. \ref{fig_8}.  Depending on the values of $\Lambda$ and $\tau$ (and depending on the history of the system in the zone of metastability), the maximum entropy state is a dipole (D), a monopole (MP) or a monopole (MN). If we fix the circulation $\Gamma$, we obtain the phase diagram in the $(\tau,E)$ plane. For $\Gamma\neq 0$, it shows the appearance of a second order phase transition in $\beta(E)$ for $\tau>\tau_c$ (for $\Gamma=0$ there is no phase transition). If we fix the energy $E$,  we obtain the phase diagram in the $(\tau,\Gamma)$ plane. As noted by Venaille \& Bouchet \cite{vb}, the point $(\Gamma=0,\tau=\tau_c)$ is a bicritical point marking the change from a first order to two second order phase transitions in $\alpha(\Gamma)$.


{\it Remark:} for illustration, we have described the phase transitions in the case of a rectangular domain and for the Laplacian operator. The generalization to an arbitrary domain and a linear operator $L$  is straightforward. In that case $\beta_{21}$ is replaced by $\beta'_1$ (the first eigenvalue of $L$ with zero mean) and  $\beta_{11}$ is replaced by $\beta''_1$ (the first eigenvalue of $L$ with non zero mean).

\section{Relaxation towards minimum enstrophy states}
\label{sec_relax}

We shall now illustrate numerically the phase transitions discussed
previously using the relaxation equations introduced in Appendix
\ref{sec_re}. These relaxation equations can serve as numerical
algorithms to compute maximum entropy states or minimum enstrophy
states with relevant constraints. Their study is also interesting in
its own right since these equations constitute non trivial dynamical
systems leading to rich bifurcations. Although these
relaxation equations do not provide a parametrization of 2D turbulence
(we have no rigorous argument for that), they may however give an idea
of the true evolution of the flow towards equilibrium. In that
respect, it would be interesting to compare these relaxation equations
with large eddy simulations (LES) of 2D turbulence. This will,
however, not be attempted in the present paper.

\subsection{Relaxation equations}

We shall numerically solve the relaxation equation of Sec. \ref{sec_relax3}.  For simplicity, we shall ignore the advective term since we are just interested in describing the bifurcations between the different equilibrium states. Then, by a proper rescaling of time, we can take $D=1$ without loss of generality.  The relaxation equation (\ref{ma10}) becomes
\begin{equation}
\label{g4}
\frac{\partial{\omega}}{\partial t}=-\left ({\omega}+\beta(t)\psi+\alpha(t)\right ),
\end{equation}
with the boundary condition $\omega=-\alpha(t)$ on the edge of the domain. The Lagrange multipliers  $\beta(t)$ and $\alpha(t)$ evolve in time according to Eqs. (\ref{ma7}) and (\ref{ma8}) in order to conserve the circulation and the energy. This leads to
\begin{eqnarray}
\label{g7}
\beta(t)=\frac{\Gamma\langle\psi\rangle-2E}{\langle\psi^2\rangle-\langle\psi\rangle^2},
\end{eqnarray}
\begin{eqnarray}
\label{g8}
\alpha(t)=-\frac{\Gamma\langle\psi^2\rangle-2E\langle\psi\rangle}{\langle\psi^2\rangle-\langle\psi\rangle^2}.
\end{eqnarray}
The rate of increase  of entropy (neg-enstrophy) is
\begin{equation}
\label{two25}
\dot S=\int  (\omega+\beta\psi+\alpha)^{2}\, d{\bf r}\ge 0.
\end{equation}
Therefore, the relaxation equation (\ref{g4}) with the constraints  (\ref{g7}) and (\ref{g8}) relaxes towards the maximum entropy state at fixed circulation and energy. Saddle points of entropy  are linearly unstable to some perturbations (in particular those described in Sec. \ref{sec_sa}).

\subsection{Geometry induced phase transitions and persistence of saddle points}
\label{sec_per}

We first consider the case of a square domain ($\tau=1<\tau_c$) and take $\Gamma=0$. For these values of parameters, the relaxation equation (\ref{g4}) admits an infinite number of steady states that are the solutions of Eq. (\ref{mes4}). However, the only stable solution is the monopole with inverse temperature $\beta_*$. It is the maximum entropy state at fixed circulation and energy. In fact, for $\Gamma=0$,  this solution is degenerate since the monopoles (MP) and (MN) have the same entropy.

\begin{figure}[h]
\center
\includegraphics[width=8cm,keepaspectratio]{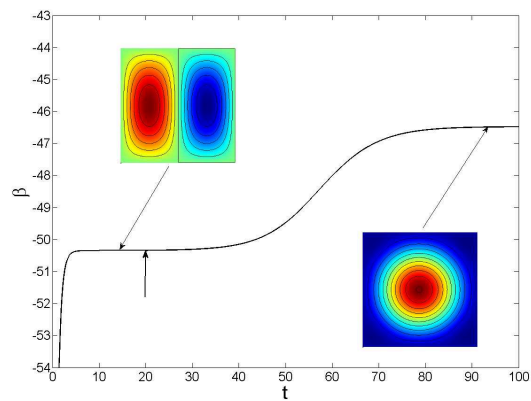}
\caption{\label{fig_6} Starting from a generic initial condition with $\Gamma=0$ in a square domain, the system relaxes towards a dipole (first plateau) although this solution is unstable (saddle point). At $t=20$ (see arrow), an optimal perturbation is applied to the dipole which quickly  destabilizes in a stable monopole (second plateau). In the absence of optimal perturbation, the system can remain frozen in the dipole for a long time.}
\end{figure}

Let us confront these theoretical results to a direct numerical simulation of Eq. (\ref{g4}). Starting from a generic initial condition (made of Gaussian peaks with positive and negative vorticity symmetrically distributed in the domain to assure $\Gamma=0$), we numerically find that the system spontaneously relaxes towards the dipole and remains in that state for a long time (see Fig.~\ref{fig_6}) although this state is predicted to be unstable (see Sec. \ref{sec_pt}). This simple numerical experiment shows that unstable states can be long-lived. In fact, the dipole is a saddle point of entropy so that it is unstable only for very specific perturbations. If these perturbations are not generated spontaneously during the relaxation process, the system can remain frozen in a saddle point for a long time. Another reason why the dipole has a long lifetime is due to the fact that the entropies of the monopole (stable) and dipole (unstable) are very close for $\Gamma=0$ since $\beta_*\approx -46.5$ and $\beta_{21}\approx -49.3$. To check that the dipole is really unstable, we have introduced {\it by hands} (see the arrow in Fig.~\ref{fig_6}) an optimal perturbation of the form $\delta\omega=1-\beta_*\phi_*$  (see Sec. \ref{sec_sa}). In that case, the dipole is immediately destabilized and the system quickly relaxes towards the monopole which is the maximum entropy state in that case.  In the case shown in Fig.~\ref{fig_6}, we obtain a monopole (MP). If we introduce an optimal perturbation with the opposite sign, we get the  monopole (MN). If we do not introduce any perturbation by hand and just let the system evolve with the numerical noise, the dipole finally destabilizes but this takes a long time (not shown) of the order $t\sim 400$.

We now consider a rectangular domain with aspect ratio $\tau=2>\tau_c$ and again take $\Gamma=0$. In that case, the maximum entropy state at fixed circulation and energy is the dipole and the monopole is unstable (saddle point).

\begin{figure}[h]
\center
\includegraphics[width=8cm,keepaspectratio]{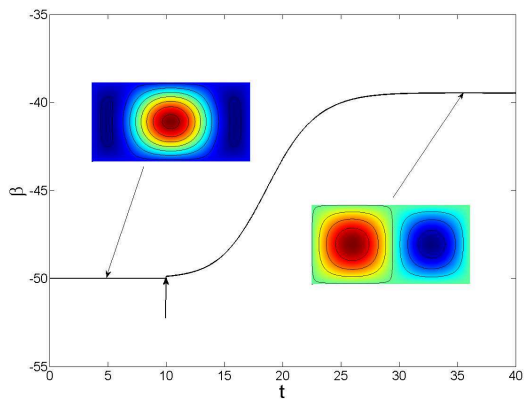}
\caption{\label{fig_7} Starting from a monopole with $\Gamma=0$ in a rectangular domain ($\tau>\tau_c$), the system remains in that state for a long time (first plateau) although this state is unstable (saddle point). At $t=10$ (see arrow), an optimal perturbation is applied to the monopole which quickly  relaxes towards a dipole (second plateau). In the absence of optimal perturbation, the system can remain frozen in the monopole state for a long time.}
\end{figure}

Starting from a generic initial condition, the system spontaneously relaxes towards the dipole and remains in this state even if very large perturbations are applied (not shown). By contrast, if we start from the monopole, we numerically observe that the system remains in that state for a very long time although this state is unstable (see Sec. \ref{sec_pt}). If we apply {\it by hands} (see the arrow in Fig.~\ref{fig_7}) an optimal perturbation of the form $\delta\omega=\psi_{21}$ (see Sec. \ref{sec_sa}), the monopole is immediately destabilized and the system quickly evolves towards the dipole (Fig.~\ref{fig_7}) which is the maximum entropy state in that case. In the absence of applied perturbation, we have not observed the destabilization of the monopole on the timescale achieved in the numerical experiment (however, if we add the advection term, the dipole is formed on a time of the order $t\sim 400$).

In conclusion, this numerical study reveals that even unstable states (saddle
points of entropy) can be naturally selected by the system and persist
for a long time. Indeed, these states are destabilized by a very
particular type of perturbations (that we call optimal) and such perturbations may
not be necessarily generated by the internal dynamics of the system. This suggests that
the system can be frozen for a long time in a quasi stationary state
(QSS) that is not necessarily a stable or metastable  steady
state of the 2D Euler equation. It can even be an unstable saddle point!
This observation has been made on the basis of the relaxation
equations that are constructed so as to relax towards a maximum entropy state. However, the same phenomenon
could appear for real flows described by the Euler or Navier-Stokes equations in numerical simulations and laboratory experiments. This could be interesting
to study in more detail.

\subsection{Metastability and hysteresis}
\label{sec_hyst}

\begin{figure}[h]
\center
\includegraphics[width=8cm,keepaspectratio,angle=0]{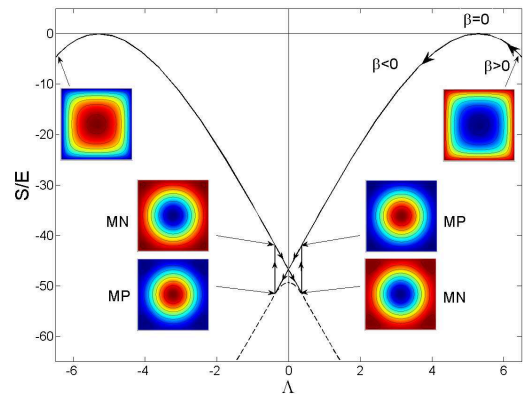}
\caption{\label{fig_1} $S/E$ ratio as a function of $\Lambda$ in a square domain.}
\end{figure}

We shall now describe the hysteretic cycle predicted by statistical mechanics (based on the neg-enstrophy) in a domain with aspect ratio $\tau<\tau_c$. In Fig. \ref{fig_1}, we plot the entropy $S/E$ as a function of the control parameter $\Lambda$. We shall assume that the energy $E$ is fixed so that $\Lambda$ basically represents the circulation $\Gamma$. The hysteresis is due to the presence of metastable states (local entropy maxima) when  $-\Lambda'_{21}<\Lambda<\Lambda'_{21}$. For $0<\Lambda<\Lambda'_{21}$, the global maximum entropy state is the direct monopole (MP) while the reversed monopole (MN) is metastable. For $-\Lambda'_{21}<\Lambda<0$, the global maximum entropy state is the direct monopole (MN) while the reversed monopole (MP) is metastable. Depending on how it has been prepared initially, the system can be found in the stable or metastable state.

\begin{figure}[h]
\center
\includegraphics[width=8cm,keepaspectratio]{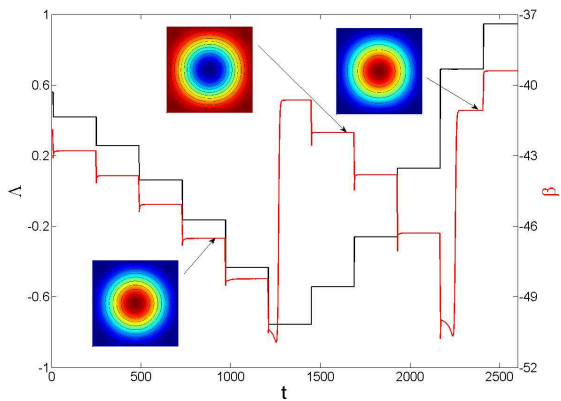}
\caption{\label{fig_9} Hysteretic cycle in a square domain,
obtained by numerical integration of Eq. (\ref{g4}). We have represented $\Lambda$ (black) and $\beta$ (red) as a function of time. Starting from a
stable state with $\Lambda\in [0;\Lambda'_{21}]$ (MP), the system is
regularly perturbed: at $t=10,250,490,730,970,1210$, we add to the
vorticity distribution the sum of a negative Gaussian peak and an
eigenmode $0.1\psi_{21}$, and let the system relax. The effect of the
Gaussian peak is to decrease $\Lambda$, while the eigenmode
destabilizes the unstable states.  For $0<\Lambda<\Lambda'_{21}$ we follow the
stable branch (MP) of Fig. \ref{fig_1} and for
$-\Lambda'_{21}<\Lambda<0$, we follow the metastable branch (MP). For
$\Lambda<-\Lambda'_{21}$, the metastable solutions (MP) no longer exist,
and the system jumps to the upper branch (MN) of Fig. \ref{fig_1}. At
$t=1450,1690,1930,2170,2410$, we add to the vorticity distribution the
sum of a positive Gaussian peak and an eigenmode $0.05\psi_{21}$. The value of
$\Lambda$ is then increased, and we follow the stable branch (MN) for
$\Lambda\in [-\Lambda'_{21};0]$ and the metastable branch (MN) for
$\Lambda\in [0;\Lambda'_{21}]$. When $\Lambda>\Lambda'_{21}$, the
metastable solutions (MN)  no longer exists, and the system jumps to the
upper branch (MP) of Fig. \ref{fig_1}.}
\end{figure}

We start from a state with large $\Lambda$ corresponding to positive temperature ($\beta>0$). In that case, the positive vorticity has the tendency to accumulate on the boundary of the domain. If we reduce $\Lambda$, we enter in the region of negative temperature states ($\beta<0$).  In that case, the positive vorticity has the tendency to accumulate at the center of the domain. For $\Lambda>0$, the (global) maximum entropy state is the monopole (MP). For $\Lambda=0$, we expect a first order phase transition from the monopole (MP) to the monopole (MN) (see Sec. \ref{sec_pt}) marked by the discontinuity of the chemical potential $\alpha$ (while $\beta$ and $S$ are still continuous). In fact, for $-\Lambda'_{21}<\Lambda\le 0$, the monopole (MP) is metastable and robust so that the system remains on this branch. Therefore, in practice, the first order phase transition does not take place. However,  for $\Lambda<-\Lambda'_{21}$, the branch of monopoles (MP) becomes unstable and the system jumps to the branch of direct monopoles (MN) which correspond to global entropy maxima. This is marked by a discontinuity of entropy (zeroth order phase transition). If we decrease $\Lambda$ sufficiently, we enter in the region of positive  temperature states ($\beta>0)$. In that case, the negative vorticity has the tendency to accumulate on the boundary of the domain. If we now increase $\Lambda$ the system follows the branch of monopoles (MN) which is stable for $\Lambda<0$ and  metastable for $0<\Lambda<\Lambda'_{21}$. Again, the first order phase transition at $\Gamma=0$ does not take place.  For $\Lambda>\Lambda'_{21}$, the branch of monopoles (MN) becomes unstable and the system jumps to the branch of direct monopoles (MP) which correspond to global entropy maxima. We have thus followed an hysteretic cycle as illustrated in Figs. \ref{fig_1} and \ref{fig_9}.

\subsection{Bifurcations in the presence of a noise}

For $\Gamma=0$ in a square domain, the monopoles (MP) and (MN) are stable and have the same entropy but remain quite distinct states (with opposite velocity). This corresponds to a parity breaking for the final organization of the system \cite{jfm}. In the presence of forcing, we expect to observe random transitions between these two solutions\footnote{This idea was initially proposed in \cite{jfm}.} similar to those observed experimentally by Sommeria \cite{sommeria} for 2D turbulence forced at small scale in a square box. Indeed, we are in a situation similar to the case of a bistable system. To observe such transitions, one possibility is to introduce a stochastic noise in the relaxation equation (\ref{g4}). Unfortunately, for a simple white noise, we did not observe any transition and we have not been able to find the properties of forcing that allow such transitions to appear. This may be due to the high entropic barriere created by the unstable (dipole) solution.  Therefore, in order to illustrate the main idea, we shall introduce a simple effective model.

\begin{figure}[h]
\center
\includegraphics[width=8cm,keepaspectratio]{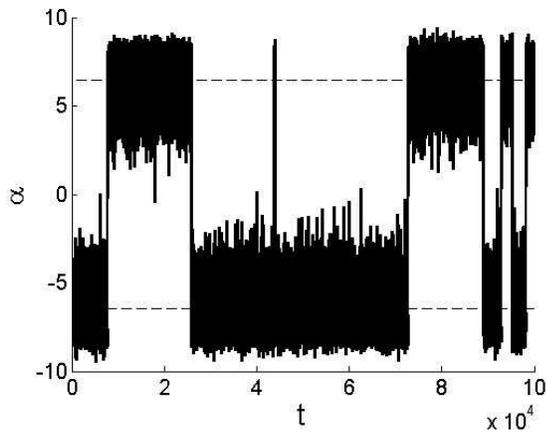}
\caption{\label{bifig}Solution of the stochastic equation (\ref{bi2}) for $\mu=1.0$ and $D=1.25$ showing random transitions between the monopoles (MP) and (MN). The dipole is always unstable.}
\end{figure}

The relevant order parameter is the chemical potential $\alpha$   which takes the values $\pm\alpha_0$ for the (stable) monopoles (MP) and (MN) and the value  $\alpha=0$ for the (unstable) dipole (see Fig. \ref{fig_4}). We shall now introduce an entropic function  $S(\alpha)$ modeled by a symmetric function with three bumps (two maxima and one minimum). Since we know the entropy (by unit of energy) of the monopoles $S_{monopoles}=\beta_*$ and the entropy of the dipole  $S_{dipole}=\beta_{21}$, we find that
\begin{equation}
\label{bi1}
S(\alpha)=(\beta_{21}-\beta_*)\left\lbrack 1-\left (\frac{\alpha}{\alpha_0}\right )^2\right\rbrack^2+\beta_*.
\end{equation}
When a forcing is present, we can propose that $\alpha$ becomes a stochastic variable described by a Langevin equation of the form
\begin{equation}
\label{bi2}
\frac{d\alpha}{dt}= \mu S'(\alpha)+\sqrt{2D} \eta(t),
\end{equation}
where $\eta(t)$ is a white noise. In the absence of forcing, Eq. (\ref{bi2}) relaxes towards one maximum of $S(\alpha)$, the monopole (MP) or the monopole (MN), and stay there permanently. In the presence of forcing, Eq. (\ref{bi2}) describes random transitions between these two states (see Fig. \ref{bifig}). This is the classical bistable system that has been studied at length in statistical mechanics and Brownian theory \cite{risken}.

Random transitions have been observed in various physical systems in
fluid mechanics (see, e.g., \cite{sommeria,benzi,bousim} and
references therein). In the present study, we have considered random
transitions between a monopole (MP) and a reversed monopole (MN). They
are associated with the first order phase transition that takes place
in a square domain when the QSS has a linear $\omega-\psi$
relationship. It would be interesting to see if they can be obtained
directly from the forced Navier-Stokes equations in situations where
the $\omega-\psi$ relationship is close to linear. Random transitions
between a unidirectional flow and a dipole have been obtained recently
by Bouchet \& Simonnet \cite{bousim} by solving numerically the forced
Navier-Stokes equations in periodic domain. However, the situation is
different (and more complex) because these two states are
characterized by different $\omega-\psi$ relationships. Indeed, the
forcing can change the shape of $\omega(\psi)$. In the situation that
we consider, the shape of $\omega(\psi)$ remains the same (linear) but
the equation $\Delta\psi=-\omega(\psi)$ determining the QSS can admit
two stable solutions (MP) and (MN). This situations is closer to that
of a bistable system and would be interesting to study numerically.

\section{Conclusion} \label{sec_conclusion}

In this paper, we have studied the maximization of the
Miller-Robert-Sommeria entropy $S_{MRS}$ at fixed energy $E$,
circulation $\Gamma$ and microscopic enstrophy $\Gamma_2^{f.g.}$ and
proved the equivalence with the minimization of the macroscopic
enstrophy $\Gamma_2^{c.g.}$ at fixed energy $E$ and circulation
$\Gamma$. This provides a justification of the minimum enstrophy
principle from statistical theory when only the microscopic enstrophy
is conserved among all the Casimir invariants. We have suggested that
relevant constraints (such as the microscopic enstrophy) are selected
by the properties of forcing and dissipation. Our simplified
thermodynamic approach leads to a mean flow characterized by a linear
$\overline{\omega}-\psi$ relationship and Gaussian fluctuations around
it. Such states can be relevant to describe certain oceanic flows
\cite{fofonoff,veronis,griffa,cummins,wang,kazantsev,niiler,marshall,bretherton,batchelor,mm,leith,k1,k2,salmon}. More
general flows with nonlinear $\overline{\omega}-\psi$ relationships
(and more general fluctuations) can be constructed in principle by
keeping other Casimir constraints in addition to
the microscopic enstrophy.

We have studied the minimization of enstrophy at fixed energy and circulation and analyzed the corresponding phase transitions with the approach of Chavanis \& Sommeria \cite{jfm}. We have discussed the link with the approach of Venaille \& Bouchet \cite{vb}. We have proposed relaxation equations to solve this minimization problem (see \cite{proc} for generalizations) and used them to illustrate the phase transitions.

One interesting result of the simulations is the observation that
saddle points of entropy can be relevant in the dynamics. Indeed,
these states are unstable only for particular perturbations that are
not necessarily generated spontaneously by the system. As a result,
they can be long-lived and robust. This observation may have
interesting application in the case of von K\'arm\'an
flows since it is found that Beltrami states are saddle points of
energy at fixed helicity, not energy minima \cite{k}. Still, it is
observed experimentally \cite{monchaux1,monchaux2} that they are
long-lived and robust.

We have also discussed in detail the metastable states that were not
considered in the study of Venaille \& Bouchet \cite{vb}. For
long-range interactions, metastable states (local entropy maxima) are
long-lived and they are as much important as fully stable states
(global entropy maxima). Interestingly, these metastable states have
negative specific heats leading to a form of ensemble inequivalence
between microcanonical and canonical ensembles (while these
ensembles are equivalent at the level of fully stable states
\cite{vb}). These metastable states can lead to an hysteresis and to
random transitions between direct monopoles and reversed
monopoles. Such transitions can also arise in more realistic fluid
systems and can have some importance in oceanography and meteorology
\cite{sommeria,benzi,bousim}.

A last remark may be in order. The MRS statistical theory of the 2D Euler equation, which is the most basic and the most rigorous, takes into account an infinite  number of constraints. When applied to real flows,
this is clearly unphysical and this leads to practical
difficulties. It has been a subject of intense debate in the last 20
years to find a practical way to deal with the constraints. Different
approaches have been proposed: some consider a point vortex
approximation where only the energy and the number of vortices in each
species matter \cite{jm}, some consider since the start only a finite
number of inviscid constraints \cite{k1,k2,salmon}, some consider a
strong mixing (or low energy) limit of the MRS statistical theory
which makes a hierarchy among the Casimir constraints \cite{jfm}, and some
model the vorticity fluctuations by a prior distribution
\cite{eht,physicaD,aussois}. In our recent works \cite{proc,cnd}, including
the present one, we have not tried to determine which approach, if
any, is the ``best''. For the moment, we just present different ways
to deal with the constraints and systematically study the
corresponding variational principles. We have also extended these
variational principles to 3D axisymmetric flows \cite{k}. These
variational principles have a long history in 2D turbulence and MHD
and one virtue of our papers is to put several variational principles
in correspondance. The determination of the ``best'' approach is still
a matter of debate and research.

\appendix

\section{Equivalence between (\ref{es12}) and (\ref{red19})}
\label{sec_eqloc}

In Sec. \ref{sec_equivgen}, we have shown the equivalence of
(\ref{es2}) and (\ref{red14}) for global maximization.  In this
Appendix, we show the equivalence of (\ref{es2}) and (\ref{red14}) for
local maximization, i.e. $\rho({\bf r},\sigma)$ is a (local) maximum
of $S[\rho]$ at fixed $E$, $\Gamma$, $\Gamma_2^{f.g.}$ and
normalization if, and only if, the corresponding coarse-grained
vorticity $\overline{\omega}({\bf r})$ is a (local) minimum of
$\Gamma_2^{c.g.}[\overline{\omega}]$ at fixed $E$ and $\Gamma$. To
that purpose, we show the equivalence between the stability criteria
(\ref{es12}) and (\ref{red19}). We use a general method similar to the
one used in \cite{cnd,nyquistgrav,cc} in related problems.

We shall determine the optimal perturbation $\delta\rho_{*}({\bf r},\sigma)$ that maximizes $\delta^{2}J[\delta\rho]$ given by Eq. (\ref{es12}) with the constraints $\delta\overline{\omega}=\int \delta\rho\sigma\, d\sigma$, $\delta\Gamma_2^{f.g.}=\int \delta\rho\sigma^2\, d\sigma d{\bf r}=0$ and $\int \delta\rho\, d\sigma=0$, where $\delta\overline{\omega}({\bf r})$ is prescribed (it is only ascribed to conserve circulation and energy at first order). Since the specification of  $\delta\overline{\omega}$ determines $\delta\psi$, hence the second integral in Eq. (\ref{es12}), we can write the variational problem in the form
\begin{eqnarray}
\label{eqa1}
\delta\left (-\frac{1}{2}\int\frac{(\delta\rho)^2}{\rho}\, d{\bf r}d\sigma\right )-\int\lambda({\bf r})\delta\left (\int\delta\rho\sigma\, d\sigma\right )\, d{\bf r}\nonumber\\
-\mu\delta\left (\int \delta\rho\sigma^2\, d\sigma d{\bf r}\right ) - \int\zeta({\bf r})\delta\left (\int\delta\rho\, d\sigma\right )\, d{\bf r}=0,\quad
\end{eqnarray}
where $\lambda({\bf r})$, $\mu$  and $\zeta({\bf r})$ are Lagrange multipliers. This gives
\begin{eqnarray}
\label{eqa2}
\delta\rho_*({\bf r},\sigma)=-\rho({\bf r},\sigma)(\mu\sigma^2+\lambda({\bf r})\sigma+\zeta({\bf r})),
\end{eqnarray}
and it is a global maximum of $\delta^{2}J[\delta\rho]$ with the previous constraints since $\delta^2(\delta^2 J)=-\int \frac{(\delta(\delta\rho))^2}{2\rho}\, d{\bf r}d\sigma< 0$ (the constraints are linear in $\delta\rho$ so their second variations vanish). The Lagrange multipliers are determined from the above-mentioned constraints. The constraints $\int \delta\rho\, d\sigma=0$ and $\delta\overline{\omega}=\int \delta\rho\sigma\, d\sigma$ lead to
\begin{eqnarray}
\zeta({\bf r})+\lambda({\bf r})\overline{\omega}({\bf r})+\mu\overline{\omega^2}({\bf r})=0,
\label{co1}
\end{eqnarray}
\begin{eqnarray}
\zeta({\bf r})\overline{\omega}({\bf r})+\lambda({\bf r})\overline{\omega^2}({\bf r})+\mu\overline{\omega^3}({\bf r})=-\delta\overline{\omega}({\bf r}).
\label{co2}
\end{eqnarray}
Now, the state $\rho({\bf r},\sigma)$ corresponds to the Gaussian distribution (\ref{es11}). Therefore, we have the well-known relations $\overline{\omega^2}({\bf r})=\overline{\omega}^2({\bf r})+\omega_2$ and $\overline{\omega^3}({\bf r})=\overline{\omega}^3({\bf r})+3\overline{\omega}({\bf r})\omega_2$ where $\omega_2=\Omega_2$ is uniform. Substituting these relations in Eqs. (\ref{co1}) and (\ref{co2}),  and solving for $\lambda({\bf r})$ and $\zeta({\bf r})$, we obtain
\begin{eqnarray}
\lambda({\bf r})=-\frac{\delta\overline{\omega}({\bf r})}{\omega_2}-2\mu\overline{\omega}({\bf r}),
\end{eqnarray}
\begin{eqnarray}
\zeta({\bf r})=\frac{\overline{\omega}({\bf r})}{\omega_2}\delta\overline{\omega}({\bf r})+\mu\overline{\omega}^2({\bf r})-\mu\omega_2.
\end{eqnarray}
Therefore, the optimal perturbation (\ref{eqa2}) can be rewritten
\begin{eqnarray}
\label{optb}
\delta\rho_{*}=-\rho\left\lbrack -\frac{\delta\overline{\omega}}{\omega_2}(\sigma-\overline{\omega})+\mu\left\lbrace (\sigma-\overline{\omega})^2-\omega_2\right\rbrace\right\rbrack.
\end{eqnarray}
The Lagrange multiplier $\mu$ is determined by substituting this expression in the constraint $\int\delta\rho\sigma^2\, d{\bf r}d\sigma=0$. Using the well-known identity $\overline{\omega^4}({\bf r})=\overline{\omega}^4({\bf r})+6\omega_2\overline{\omega}^2({\bf r})+3\omega_2^2$  valid for a Gaussian distribution, we obtain after some simplifications
\begin{eqnarray}
\label{mu}
\mu=\frac{\int\overline{\omega}\delta\overline{\omega}\, d{\bf r}}{\omega_2^2}.
\end{eqnarray}
Therefore, the optimal perturbation (\ref{eqa2}) is given by
Eq. (\ref{optb}) with Eq. (\ref{mu}). Since this perturbations
maximizes $\delta^{2}J[\delta\rho]$ with the above-mentioned
constraints, we have $\delta^{2}J[\delta\rho]\le
\delta^{2}J[\delta\rho_*]$. Explicating $\delta^{2}J[\delta\rho_*]$
using Eqs. (\ref{optb}) and (\ref{mu}), we obtain after simple
calculations
\begin{eqnarray}
\label{eqa3b}
\delta^{2}J[\delta\rho]\le -\frac{1}{2\omega_2}\int (\delta\overline{\omega})^2\, d{\bf r}-\frac{1}{\omega_2^2}\left (\int\overline{\omega}\delta\overline{\omega}\, d{\bf r}\right )^2\nonumber\\
-\frac{1}{2}\beta\int\delta\overline{\omega}\delta\psi\, d{\bf r}.
\end{eqnarray}
The r.h.s. returns the functional appearing in Eq. (\ref{ev13}). We
have already explained in Sec. \ref{sec_equiv} that for the class of
perturbations that we consider ($\delta\Gamma=\delta E=0$) the second
integral vanishes. Therefore, the foregoing inequality can be
rewritten
\begin{eqnarray}
\label{eqa4}
\delta^{2}J[\delta\rho]\le -\frac{1}{2\omega_2}\int (\delta\overline{\omega})^2\, d{\bf r}-\frac{1}{2}\beta\int\delta\overline{\omega}\delta\psi\, d{\bf r},
\end{eqnarray}
where the r.h.s. is precisely the functional appearing in
Eq. (\ref{red19}). Furthermore, there is equality in Eq. (\ref{eqa4})
iff $\delta\rho=\delta\rho_*$. This proves that the stability criteria
(\ref{es12}) and (\ref{red19}) are equivalent. Indeed: (i) if
inequality (\ref{red19}) is fulfilled for all perturbations
$\delta\overline{\omega}$ that conserves circulation and energy at
first order, then according to Eq. (\ref{eqa4}), we know that
inequality (\ref{es12}) is fulfilled for all perturbation $\delta\rho$
that conserves circulation, energy, fine-grained enstrophy and
normalization at first order; (ii) if there exists a perturbation
$\delta\overline{\omega}_*$ that makes
$\delta^{2}J[\delta\overline{\omega}]>0$, then the perturbation
$\delta\rho_*$ given by Eq. (\ref{optb}) with Eq. (\ref{mu}) and
$\delta\overline{\omega}=\delta\overline{\omega}_*$ makes
$\delta^{2}J[\delta\rho]>0$. In conclusion, the stability criteria (\ref{es12}) and (\ref{red19}) are equivalent.

\section{Eigenvalues and eigenfunctions of the Laplacian in a rectangular domain}
\label{sec_eigen}

We define the eigenfunctions and eigenvalues of the Laplacian by
\begin{eqnarray}
\label{ei1}
\Delta\psi_n=\beta_n\psi_n,
\end{eqnarray}
with $\psi_n=0$ on the domain boundary. These eigenfunctions are orthogonal and normalized so that $\langle\psi_n\psi_m\rangle=\delta_{nm}$. Since $-\int (\nabla\psi_n)^2\, d{\bf r}=\beta_n\int \psi_n^2\, d{\bf r}$, we note that $\beta_n<0$. Following Chavanis \& Sommeria \cite{jfm}, we distinguish two types of eigenmodes: the odd eigenmodes $\psi_n'$ such that $\langle \psi_n'\rangle=0$ and the even eigenmodes $\psi_n''$ such that $\langle \psi_n''\rangle\neq 0$. We note $\beta_n'$ and $\beta_n''$ the corresponding eigenvalues.

In a rectangular domain of unit area whose sides are denoted $a=\sqrt{\tau}$ and $b=1/\sqrt{\tau}$ (where $\tau=a/b$ is the aspect ratio), the eigenmodes and eigenvalues are
\begin{eqnarray}
\label{ei5}
\psi_{mn}=2\sin(m\pi x/\sqrt{\tau})\sin(n\pi\sqrt{\tau}y),
\end{eqnarray}
\begin{eqnarray}
\label{ei6}
\beta_{mn}=-\pi^2\left (\frac{m^2}{\tau}+\tau n^2\right ),
\end{eqnarray}
where the origin of the Cartesian frame is taken at the lower left corner of the domain. The integer $m\ge 1$ gives the number of vortices along the $x$-axis and $n\ge 1$ the number of vortices along the $y$-axis. We have $\langle \psi_{mn}\rangle =0$ if $m$ or $n$ is even and  $\langle \psi_{mn}\rangle \neq 0$ if $m$ and $n$ are odd.

The differential equation (\ref{c2}) can be solved analytically by decomposing the field $\phi$ on the eigenmodes as $\phi=\sum_{mn} c_{mn}\psi_{mn}$ and using the identity $1=\sum_{mn} \langle \psi_{mn}\rangle \psi_{mn}$. This yields Eq. (\ref{c3}) from which we obtain
\begin{eqnarray}
\label{ei2}
\langle\phi\rangle=\sum_{mn} \frac{\langle\psi_{mn}\rangle^2}{\beta-\beta_{mn}},
\end{eqnarray}
\begin{eqnarray}
\label{ei3}
\langle\phi^2\rangle=\sum_{mn} \frac{\langle\psi_{mn}\rangle^2}{(\beta-\beta_{mn})^2}=-\frac{d\langle\phi\rangle}{d\beta}.
\end{eqnarray}
We note in particular that
\begin{eqnarray}
\label{ei4}
\langle\phi\rangle-\beta\langle\phi^2\rangle=-\sum_{mn} \frac{\beta_{mn}\langle\psi_{mn}\rangle^2}{(\beta-\beta_{mn})^2}>0.
\end{eqnarray}

\section{Temporal evolution of the different modes}

The relaxation equation (\ref{g4}) can be solved analytically by decomposing the vorticity and the stream function on the eigenmodes of the Laplacian. Using the Poisson equation, we get $\omega({\bf r},t)=\sum_{n}a_n(t)\psi_n({\bf r})$ and $\psi({\bf r},t)=\sum_{n}b_n(t)\psi_n({\bf r})$ with  $b_n(t)=-a_n(t)/\beta_n$. Substituting these expressions in Eq. (\ref{g4}) and using the identity $1=\sum_n \langle\psi_n\rangle\psi_n$, we obtain the ordinary differential equations
\begin{eqnarray}
\label{e3}
\frac{d a_n}{dt}+\left (1-\frac{\beta(t)}{\beta_n}\right )a_n=-\alpha(t)\langle\psi_n\rangle,
\end{eqnarray}
for all $n$. The evolution of the Lagrange multipliers is given by Eqs. (\ref{g7}) and (\ref{g8}) with $\langle\psi\rangle=\sum_n b_n(t)\langle\psi_n\rangle$ and $\langle\psi^2\rangle=\sum_n b_n^2(t)$. The modes are coupled through the Lagrange multipliers in order to assure the conservation of energy and circulation.

In the grand canonical description in which $\beta$ and $\alpha$ are constants, the foregoing differential equation can be integrated straightforwardly, yielding
\begin{eqnarray}
\label{e3gc}
a_n(t)=\left (a_n(0)+\frac{\alpha\langle\psi_n\rangle}{1-\beta/\beta_n}\right )e^{-(1-\beta/\beta_n)t}-\frac{\alpha\langle\psi_n\rangle}{1-\beta/\beta_n}. \nonumber\\
\end{eqnarray}
In that case, a steady state of the relaxation equation is stable iff $\beta>\beta'_1$  where $\beta'_1$ is the largest eigenvalue of the Laplacian. The condition $\beta>\beta'_1$ is a necessary and sufficient condition for the steady state to be a global maximum of the grand potential $G=S-\beta E-\alpha\Gamma$. That functional is related to the Arnol'd energy-Casimir functional used to settle the nonlinear dynamical stability of a steady state of the 2D Euler equation \cite{proc}.

\section{Relaxation equations}
\label{sec_re}

\subsection{Relaxation equations associated with the maximization problem (\ref{es2})}
\label{sec_relax1}

In this Appendix, we construct relaxation equations associated with
the maximization problem (\ref{es2}) corresponding to the
energy-enstrophy-circulation statistical theory. These relaxation equations can
serve as a numerical algorithm to solve this constrained maximization
problem.  In the past, Robert \& Sommeria \cite{rsmepp} have proposed
relaxation equations that conserve all the Casimirs and increase the
entropy. Here, we use a different approach because we want to conserve
only the microscopic enstrophy (not all the Casimirs). Thus, the form
of the relaxation equations will be different. In particular, they
will involve a current in the space of vorticity levels $\sigma$
\cite{proc,cnd} instead of a current in the space of positions
\cite{rsmepp}.

We construct a set of relaxation equations that increase $S[\rho]$ while
conserving $E$, $\Gamma$ and
$\Gamma_2^{f.g.}$ using a Maximum Entropy Production Principle.  The
dynamical equation that we consider can be written as
\begin{eqnarray}
\frac{\partial\rho}{\partial t}+{\bf u}\cdot \nabla\rho=-\frac{\partial J}{\partial\sigma},\label{re1}
\end{eqnarray}
where $J$ is an unknown current to be chosen so as to increase $S[\rho]$ while conserving the constraints. The local normalization $\int \rho d\sigma=1$ is satisfied provided that $J \rightarrow 0$ as $\sigma \rightarrow \pm \infty$. Multiplying Eq. (\ref{re1})  by $\sigma$ and integrating over the levels, we get
\begin{equation}
\frac{\partial\overline{\omega}}{\partial t}+{\bf u}\cdot \nabla\overline{\omega}=\int J d\sigma\equiv X. \label{re2}
\end{equation}
Next, multiplying Eq. (\ref{re1})  by $\sigma^2$ and integrating over the levels, we obtain
\begin{equation}
\frac{\partial\overline{\omega^2}}{\partial t}+{\bf u}\cdot \nabla\overline{\omega^2}=2\int J\sigma d\sigma. \label{re3}
\end{equation}
From Eqs. (\ref{re2}) and (\ref{re3}), we find that
\begin{equation}
\frac{\partial {\omega_2}}{\partial t}+{\bf u}\cdot \nabla {\omega_2}=2\int J(\sigma-\overline{\omega}) d\sigma. \label{re4}
\end{equation}
Using Eq. (\ref{re1}), the time variations of $S[\rho]$ are given by
\begin{equation}
\dot{S}=-\int \frac{J}{\rho} \frac{\partial\rho}{\partial\sigma} \, d{\bf r}d\sigma,\label{re5}
\end{equation}
and  the time variations of $E,\Gamma,{\Gamma}_2^{f.g.}$ are given by
\begin{equation}
\dot{E}= \int J\psi \, d{\bf r}d\sigma=0, \label{re6a}
\end{equation}
\begin{equation}
\dot{\Gamma}=\int J \, d{\bf r}d\sigma=0,\\\label{re6b}
\end{equation}
\begin{equation}
\dot{\Gamma}_2^{f.g.}=2\int J \sigma \, d{\bf r}d\sigma=0.\label{re6c}
\end{equation}
Following the Maximum Entropy Production Principle, we maximize $\dot{S}$ with $\dot{{E}}=\dot{{\Gamma}}=\dot{\Gamma}_2^{f.g.}=0$ and the additional constraint
\begin{equation}
\int \frac{J^2}{2\rho} \, d\sigma\le C({\bf r},t),\label{re7}
\end{equation}
putting some physical bound on the diffusion current.  The
variational principle can be written in the form
\begin{eqnarray}
\delta \dot{S} - \beta(t) \delta \dot{{E}} - \alpha(t) \delta \dot{\Gamma} - \alpha_2(t) \delta \dot{\Gamma}_2^{f.g.} & - & \nonumber \\
\int \frac{1}{D({\bf r},t)} \delta \left( \int \frac{J^2}{2\rho} d\sigma \right) d{\bf r} =0,\label{re8}
\end{eqnarray}
where $\beta(t)$, $\alpha(t)$, $\alpha_2(t)$ and $D({\bf r},t)$ are time dependent Lagrange multipliers associated with the constraints.
This leads to  the following optimal current
\begin{eqnarray}
J=-D\left\lbrack \frac{\partial\rho}{\partial \sigma}+\rho\left (\beta(t)\psi +\alpha(t)+2\alpha_2(t) \sigma \right )\right\rbrack. \label{re9}
\end{eqnarray}
Therefore, the relaxation equation for the vorticity distribution is
\begin{eqnarray}
\frac{\partial\rho}{\partial t}&+&{\bf u}\cdot \nabla\rho\nonumber\\
&=&\frac{\partial}{\partial\sigma}\left\lbrace D\left\lbrack \frac{\partial\rho}{\partial \sigma}+\rho\left (\beta(t)\psi +\alpha(t)+2\alpha_2(t) \sigma \right )\right\rbrack\right\rbrace.\qquad\label{re15}
\end{eqnarray}
Integrating Eq. (\ref{re9}) over $\sigma$, we obtain
\begin{eqnarray}
X=-D \left (\beta(t)\psi+\alpha(t)+2\alpha_{2}(t)\overline{\omega}\right).\label{re10}
\end{eqnarray}
Inserting Eq. (\ref{re10}) into Eq.~(\ref{re2}) leads to the
following relaxation equation for the mean flow
\begin{eqnarray}
\frac{\partial\overline{\omega}}{\partial t}+{\bf u}\cdot \nabla\overline{\omega}=-D \left (\beta(t)\psi+\alpha(t)+2\alpha_{2}(t)\overline{\omega}\right).\label{re16}
\end{eqnarray}
For the boundary condition, we shall take $\beta(t)\psi+\alpha(t)+2\alpha_{2}(t)\overline{\omega}=0$ on the domain boundary so as to be consistent with the equilibrium state where this quantity vanishes in the whole domain. Since $\psi=0$ on the boundary, we finally get $\overline{\omega}=-\alpha(t)/(2\alpha_2(t))$ on the domain boundary.
A relaxation equation can also be written for the centered variance $\omega_2$. Using Eqs. (\ref{re9}) and (\ref{re4}), we obtain
\begin{equation}
\frac{\partial\omega_2}{\partial t}+{\bf u}\cdot \nabla \omega_2=2D\left (1-2\alpha_2(t)\omega_2\right ). \label{re17}
\end{equation}
Finally, in Eqs. (\ref{re15}), (\ref{re16}) and (\ref{re17}), the
Lagrange multipliers evolve so as to satisfy the
constraints. Substituting Eq. (\ref{re9}) in Eqs. (\ref{re6a}), (\ref{re6b}) and (\ref{re6c}), we
obtain the algebraic equations
\begin{eqnarray}
\langle\psi^2\rangle\beta(t)+\langle\psi\rangle\alpha(t)+4E\alpha_2(t)=0,\label{re11}
\end{eqnarray}
\begin{eqnarray}
\langle\psi\rangle\beta(t)+\alpha(t)+2\Gamma\alpha_2(t)=0,\label{re12}
\end{eqnarray}
\begin{eqnarray}
2E\beta(t)+\Gamma\alpha(t)+2\Gamma_2^{f.g.}\alpha_2(t)=1.\label{re13}
\end{eqnarray}
where $\langle X\rangle=\int X\, d{\bf r}$. Substituting $\partial\rho/\partial\sigma$ taken from Eq. (\ref{re9}) in Eq. (\ref{re5}) and using the constraints   (\ref{re6a})-(\ref{re6c}), we easily obtain
\begin{eqnarray}
\dot S=\int \frac{J^2}{D\rho}\, d{\bf r}d\sigma,\label{re14}
\end{eqnarray}
so that $\dot S\ge 0$ provided that $D$ is positive. On the other hand
$\dot S=0$ iff $J=0$ leading to the Gibbs state (\ref{es7}). From
Lyapunov's direct method, we conclude that these relaxation equations
tend to a maximum of entropy at fixed energy, circulation and
microscopic enstrophy. Note that during the relaxation process, the
distribution of vorticity is not Gaussian but changes with time
according to Eq. (\ref{re15}). The vorticity distribution is Gaussian
only at equilibrium. Therefore, these relaxation equations describe
not only the evolution of the mean flow according to Eq. (\ref{re16})
but also the evolution of the full vorticity distribution according to
Eq. (\ref{re15}). We stress, however, that these equations are purely
phenomenological and that there is no compelling reason why they
should give an accurate description of the real dynamics. However,
they can be used at least as a numerical algorithm to compute the
statistical equilibrium state. Indeed, these equations can only relax
towards an entropy maximum at fixed energy, circulation and
microscopic enstrophy, not towards a minimum or a saddle point that
are linearly unstable with respect to these equations\footnote{In
fact, it is shown in Sec. \ref{sec_relax} that the system can remain
blocked in an unstable state (saddle point of entropy) if the dynamics
does not spontaneously develop the ``dangerous'' perturbations that
make it unstable. This is because the system is
unstable for {\it some} perturbations but not for {\it all}
perturbations. Therefore, we must keep in mind this property when we
use the relaxation equations.}.

\subsection{Relaxation equations associated with the maximization problem (\ref{ev6})}
\label{sec_relax2}

We shall now introduce a set of relaxation equations associated with
the maximization problem (\ref{ev6}). We write the dynamical equation
as
\begin{eqnarray}
\frac{\partial\overline{\omega}}{\partial t}+{\bf u}\cdot \nabla\overline{\omega}=X, \label{res1}
\end{eqnarray}
where $X$ is an  unknown quantity to be chosen so as to increase $S[\overline{\omega}]$ while conserving $E$, $\Gamma$ and $\Gamma_2^{f.g.}$.
The time variations of $S$ are given by
\begin{equation}
\dot{S}=-\frac{1}{\Omega_2(t)}\int \overline{\omega} X\, d{\bf r},\label{res2}
\end{equation}
where $\Omega_2(t)$ is determined by the constraint on microscopic enstrophy leading to
\begin{equation}
\Omega_{2}(t)=\Gamma_2^{f.g.}-\int \overline{\omega}^2 \, d{\bf r},
\label{res3}
\end{equation}
at each time. On the other hand, the time variations of $E$ and $\Gamma$ are
\begin{eqnarray}
\dot{E}=\int X\psi \, d{\bf r}=0,\label{res3b}\\
\dot{\Gamma}=\int X \, d{\bf r}=0.\label{res4}
\end{eqnarray}

Following the Maximum Entropy Production Principle, we maximize $\dot{S}$ with $\dot{E}=\dot{\Gamma}=0$ (the conservation of microscopic enstrophy has been taken into account in Eq. (\ref{res3})) and the additional constraint
\begin{equation}
\frac{X^2}{2}\le C({\bf r},t).\label{res5}
\end{equation}
The variational principle can be written in the form
\begin{eqnarray}\label{res6}
\delta \dot{S} - \beta(t) \delta \dot{E} - \alpha(t) \delta \dot{\Gamma} -
\int \frac{1}{D({\bf r},t)} \delta \left(\frac{X^2}{2} \right) d{\bf r}=0,\nonumber\\
\end{eqnarray}
and it leads to the optimal quantity
\begin{eqnarray}
X=-D \left( \beta(t) {\psi} +\alpha(t)+\frac{1}{\Omega_2(t)}\overline{\omega} \right).\label{res7}
\end{eqnarray}
Inserting Eq. (\ref{res7}) in Eq. (\ref{res1}), we obtain
\begin{eqnarray}
\frac{\partial\overline{\omega}}{\partial t}+{\bf u}\cdot \nabla\overline{\omega}=-D \left( \beta(t) {\psi} +\alpha(t)+\frac{1}{\Omega_2(t)}\overline{\omega} \right),\label{res10}
\end{eqnarray}
with $\overline{\omega}=-\alpha(t)\Omega_2(t)$ on the domain boundary.
The Lagrange multipliers evolve so as to satisfy the constraints. Substituting Eq. (\ref{res7}) in Eqs. (\ref{res3b})-(\ref{res4}) and recalling Eq. (\ref{res3}), we obtain the algebraic equations
\begin{equation}
\Omega_{2}(t)=\Gamma_2^{f.g.}-\int \overline{\omega}^2 \, d{\bf r},
\label{res3c}
\end{equation}
\begin{eqnarray}
\langle\psi^2\rangle\beta(t)+\langle\psi\rangle\alpha(t)=-\frac{2E}{\Omega_2(t)},\label{res7b}
\end{eqnarray}
\begin{eqnarray}
\langle\psi\rangle\beta(t)+\alpha(t)=-\frac{\Gamma}{\Omega_2(t)}.\label{res8}
\end{eqnarray}
Substituting $\overline{\omega}$ taken from Eq. (\ref{res7}) in Eq. (\ref{res2}) and using the constraints (\ref{res3b})-(\ref{res4}), we easily obtain
\begin{eqnarray}
\dot S=\int \frac{X^2}{D}\, d{\bf r},\label{res9}
\end{eqnarray}
so that $\dot S\ge 0$ provided that $D$ is positive. On the other hand
$\dot S=0$ iff $X=0$ leading to the condition of
equilibrium (\ref{ev12}). From Lyapunov's direct method, we conclude
that these relaxation equations tend to a maximum of entropy at fixed
energy, circulation and microscopic enstrophy.

The relaxation equation (\ref{res10}) is similar to Eq. (\ref{re16}) but
the constraints determining the evolution of the Lagrange multipliers
are different. More precisely, Eqs. (\ref{res7b}) and (\ref{res8}) are equivalent to Eqs. (\ref{re12}) and (\ref{re13}) but Eq. (\ref{re11}) has been replaced by Eq. (\ref{res3c}). Indeed, in the present approach, the vorticity
distribution is always Gaussian during the dynamical evolution. It is
given by Eq. (\ref{ev2}) at any time, i.e.
\begin{equation}
\rho({\bf r},\sigma,t) = \frac{1}{\sqrt{2\pi\Omega_2(t)}}e^{-\frac{(\sigma-\overline{\omega}({\bf r},t))^2}{2\Omega_2(t)}}.
\label{ev2n}
\end{equation}
By contrast, in the approach of Sec. \ref{sec_relax1}, the vorticity
distribution changes with time. Therefore, the dynamical evolution is
different. However, in the two approaches, the equilibrium state is
the same, i.e. it solves the maximization problem (\ref{es2}). This is
sufficient if we use these relaxation equations as numerical
algorithms to compute the maximum entropy state.

{\it Remark:} Using Eqs. (\ref{res1})-(\ref{res2}), it is easy to show that $\dot\Gamma_2^{c.g.}=-2\Omega_2(t)\dot S$  so that $\dot\Gamma_2^{c.g.}\le 0$ since $\Omega_2(t)\ge 0$ (by Schwartz inequality). Therefore, the macroscopic enstrophy decreases monotonically through the relaxation equations. This is to be expected since the maximization problem (\ref{ev6}) is equivalent to the minimization of the macroscopic enstrophy at fixed energy and circulation (see Sec. \ref{sec_reduced}).

{\it Alternative relaxation equation:} writing the r.h.s. of Eq. (\ref{res1}) in the form of the divergence of a current in order to conserve the circulation, and using a MEPP, we obtain a relaxation of the form \cite{proc}:
\begin{eqnarray}
\frac{\partial\overline{\omega}}{\partial t}+{\bf u}\cdot \nabla\overline{\omega}=\nabla\cdot \left\lbrack D \left( \frac{1}{\Omega_2(t)}\nabla\overline{\omega}+\beta(t) \nabla{\psi} \right)\right \rbrack,\label{rey1}
\end{eqnarray}
where $\Omega_2(t)$ is given by Eq. (\ref{res3c}) and $\beta(t)$ by
\begin{eqnarray}
\beta(t)=\frac{-\int D\nabla\overline{\omega}\cdot\nabla\psi\, d{\bf r}}{\Omega_2(t)\int D(\nabla\psi)^2\, d{\bf r}}.\label{rey2}
\end{eqnarray}
The boundary conditions are $(\frac{1}{\Omega_2(t)}\nabla\overline{\omega}+\beta(t) \nabla{\psi})\cdot {\bf n}=0$ on the domain boundary. This relaxation equation satisfies the same general properties as Eq. (\ref{res10}).

\subsection{Relaxation equations associated with the
maximization problem (\ref{red14})}
\label{sec_relax3}

We shall introduce a set of relaxation equations associated with the
maximization problem (\ref{red14}). We write the dynamical equation as
\begin{eqnarray}
\frac{\partial\overline{\omega}}{\partial t}+{\bf u}\cdot \nabla\overline{\omega}=X, \label{ma1}
\end{eqnarray}
where $X$ is an  unknown quantity to be chosen so as to increase $S[\overline{\omega}]$ while conserving $E$ and $\Gamma$.
The time variations of $S$ are given by
\begin{equation}
\dot{S}=-\int \overline{\omega} X\, d{\bf r}. \label{ma2}
\end{equation}
On the other hand, the time variations of $E$ and $\Gamma$ are
\begin{eqnarray}
\dot{E}=\int X\psi \, d{\bf r}=0,\label{ma2b}\\
\dot{\Gamma}=\int X \, d{\bf r}=0. \label{ma3}
\end{eqnarray}

Following the Maximum Entropy Production Principle, we maximize $\dot{S}$ with $\dot{E}=\dot{\Gamma}=0$ and the additional constraint
\begin{equation}
\frac{X^2}{2}\le C({\bf r},t). \label{ma4}
\end{equation}
The variational principle can be written in the form
\begin{eqnarray} \label{ma5}
\delta \dot{S} - \beta(t) \delta \dot{E} - \alpha(t) \delta \dot{\Gamma} -
\int \frac{1}{D({\bf r},t)} \delta \left(\frac{X^2}{2} \right) d{\bf r}=0,\nonumber\\
\end{eqnarray}
and we obtain
\begin{eqnarray}
X=-D \left( \beta(t) {\psi} +\alpha(t)+\overline{\omega} \right). \label{ma6}
\end{eqnarray}
Substituting Eq. (\ref{ma6}) in Eq. (\ref{ma1}), we obtain
\begin{eqnarray}
\frac{\partial\overline{\omega}}{\partial t}+{\bf u}\cdot \nabla\overline{\omega}=-D \left( \beta(t) {\psi} +\alpha(t)+\overline{\omega} \right), \label{ma10}
\end{eqnarray}
with $\overline{\omega}=-\alpha(t)$ on the domain boundary. The Lagrange multipliers $\beta(t)$ and $\alpha(t)$ evolve so as to satisfy the constraints. Substituting Eq. (\ref{ma6}) in Eqs. (\ref{ma2b}) and (\ref{ma3}), we obtain
the algebraic equations
\begin{eqnarray}
\langle\psi^2\rangle\beta(t)+\langle\psi\rangle\alpha(t)=-2E, \label{ma7}
\end{eqnarray}
\begin{eqnarray}
\langle\psi\rangle\beta(t)+\alpha(t)=-{\Gamma}. \label{ma8}
\end{eqnarray}
Substituting $\overline{\omega}$ taken from Eq. (\ref{ma6}) in Eq. (\ref{ma2}) and using the constraints (\ref{ma2b})-(\ref{ma3}), we  easily obtain
\begin{eqnarray}
\dot S=\int \frac{X^2}{D}\, d{\bf r}, \label{ma9}
\end{eqnarray}
so that $\dot S\ge 0$ provided that $D$ is positive. On the other hand $\dot S=0$ iff $X=0$ leading to the condition of equilibrium  (\ref{red18}). From Lyapunov's direct method, we conclude that these relaxation equations tend to a maximum of entropy (or a minimum of enstrophy) at fixed energy and circulation.

{\it Alternative relaxation equation:} writing the r.h.s. of Eq. (\ref{ma1}) in the form of the divergence of a current in order to conserve the circulation, and using a MEPP, we obtain a relaxation of the form \cite{proc}:
\begin{eqnarray}
\frac{\partial\overline{\omega}}{\partial t}+{\bf u}\cdot \nabla\overline{\omega}=\nabla\cdot \left\lbrack D \left(\nabla\overline{\omega}+\beta(t) \nabla{\psi} \right)\right \rbrack,\label{rey1b}
\end{eqnarray}
\begin{eqnarray}
\beta(t)=\frac{-\int D\nabla\overline{\omega}\cdot\nabla\psi\, d{\bf r}}{\int D(\nabla\psi)^2\, d{\bf r}}.\label{rey2b}
\end{eqnarray}
The boundary conditions are $(\nabla\overline{\omega}+\beta(t) \nabla{\psi})\cdot {\bf n}=0$ on the domain boundary. This relaxation equation satisfies the same general properties as Eq. (\ref{ma10}). If we assume that $D$ is constant, the foregoing equation can be rewritten
\begin{eqnarray}
\frac{\partial\overline{\omega}}{\partial t}+{\bf u}\cdot \nabla\overline{\omega}=D \left(\Delta\overline{\omega}-\beta(t) \overline{\omega} \right),\label{rey1bis}
\end{eqnarray}
\begin{eqnarray}
\beta(t)=-\frac{\int \overline{\omega}^2\, d{\bf r}}{2 E}=\frac{S(t)}{E},\label{rey2bis}
\end{eqnarray}
where we have used an integration by parts to obtain the second term of  Eq. (\ref{rey2bis}).

{\it Remark 1}: Since the relaxation equations derived in this section solve (\ref{ens1}), they can also be used as a numerical algorithm to construct nonlinearly dynamically stable stationary solutions of the 2D Euler equations characterized by a linear $\omega-\psi$ relationship (see Secs. \ref{sec_ss} and \ref{sec_ens}) independently of the statistical mechanics interpretation.

{\it Remark 2}: Since the EHT thermodynamical equilibrium  with a Gaussian prior is equivalent to (\ref{red14}), the relaxation equations derived in this section coincide with a particular case of the relaxation equations derived in \cite{cnd} (the ones corresponding to a Gaussian prior).

{\it Remark 3}: Since the optimization problems (\ref{es2}), (\ref{ev6}) and (\ref{red14}) are equivalent, the corresponding relaxation equations derived in Appendices  \ref{sec_relax1}, \ref{sec_relax2} and \ref{sec_relax3} have the same equilibrium states. However, the dynamics leading to these equilibrium states is different in each case because the constraints are different.

\end{document}